\documentclass{ecai}
\usepackage{graphicx}
\usepackage{latexsym}
\usepackage{amsmath}
\usepackage{algorithm} % http://ctan.org/pkg/algorithms
\usepackage{algpseudocode} % http://ctan.org/pkg/algorithmicx
\usepackage{amssymb}
\usepackage{caption}
\usepackage{float}

\ecaisubmission   % inserts page numbers. Use only for submission of paper.
\begin{document}

\begin{frontmatter}

\title{Approximate Nash Equilibrium Learning for n-Player Markov Games in Dynamic Pricing}

% \author[A]{\fnms{Larkin Liu}~\snm{}\orcid{0000-0002-9375-1035}\thanks{Corresponding Author. Email: larkin.liu@tum.de}}
\author{Larkin Liu}

% \author[A]{\fnms{Larkin Liu}~\snm{}\orcid{0000-0002-9375-1035}\thanks{Corresponding Author. Email: larkin.liu@tum.de}}

% \author{
%     Larkin Liu \\
%     Technical University of Munich \\
%     \texttt{larkin.liu@tum.de}
% }

% \author[B]{\fnms{Second}~\snm{Author}\orcid{....-....-....-....}}
% \author[B]{\fnms{Third}~\snm{Author}\orcid{....-....-....-....}} % use of \orcid{} is optional

% \author[A]{\fnms{Larkin Liu}\thanks{Corresponding Author. Email: larkin.liu@tum.de}}
% \author{\fnms{Larkin Liu}}

\address{Technical University of Munich \\ \texttt{larkin.liu@tum.de}}
% \address[B]{Munich Data Science Institute}

\begin{abstract}
We investigate Nash equilibrium learning in a competitive Markov Game (MG) environment, where multiple agents compete, and multiple Nash equilibria can exist. In particular, for an oligopolistic dynamic pricing environment, exact Nash equilibria are difficult to obtain due to the curse-of-dimensionality. We develop a new model-free method to find approximate Nash equilibria. Gradient-free black box optimization is then applied to estimate $\epsilon$, the maximum reward advantage of an agent unilaterally deviating from any joint policy, and to also estimate the $\epsilon$-minimizing policy for any given state. The policy-$\epsilon$ correspondence and the state to $\epsilon$-minimizing policy are represented by neural networks, the latter being the Nash Policy Net. During batch update, we perform Nash Q learning on the system, by adjusting the action probabilities using the Nash Policy Net. We demonstrate that an approximate Nash equilibrium can be learned, particularly in the dynamic pricing domain where exact solutions are often intractable.
\end{abstract}

\end{frontmatter}

\section{Introduction}
%The traditional page limit for ECAI long papers is {\bf 7 (six)} pages
%in the required format. The traditional page limit for short
%submissions is {\bf 2} pages.
%
%However, these page limits may change from one ECAI to
%another. Consult the most recent Call For Papers (CFP) for the most
%up-to-date page limits.

The application of deep reinforcement learning to solve single player Markov Decision Processes are relatively well-researched in today's machine learning literature, however, the application of novel deep reinforcement learning methods to solve multi-agent competitive MDP's are relatively few. Particularly, the challenge revolves around competing objectives and solving for, often computationally intractable, Nash equilibria \cite{Nash:1950} in a competitive setting, where agents cannot merely maximize only their respective Q functions. Nash equilibria in Markov Games are particularly useful in the area of \textit{dynamic pricing}, a well-known problem in today's online eCommerce marketplaces.  In a \textit{dynamic pricing} game, multiple firms compete for market share of their product on the same online platform. To model this market, we adopt a modified version of Bertrand oligopoly \cite{bertrand:1883}, where the marginal cost of production is 0, and where purchasing behaviour is probabilistic.

% where multiple firms compete to maximize their revenue, and consumers tend to purchase the item of the lowest price, ceteris paribus. 

% The prices set by the firms, also constituting the actions, affect the demand of the product and consequently the profits in the current time and into the future. We adopt the model of Bertrand oligopoly to model this process, where multiple firms compete to maximize their revenue, and consumers tend to purchase the item of the lowest price, ceteris paribus. However, the traditional economic theory of Bertrand competition \cite{bertrand:1883} does not hold in the era of large scale data, where many factors contribute to the the variation of demand (i.e. switching costs, advertising, brand loyalty) etc. 

\subsection{Market Oligopoly}

Especially in today's online commerce industries, dynamic pricing provides platform firms with an adversarial advantage and increases profit margins in a competitive economy. In an ideal case, a firm's pricing strategy should consider the actions of other firms, providing the justification for Nash equilibrium policies over unilateral reward maximization policies. In previous approaches, \cite{Liu:2019-dyp} modelled the dynamic pricing problem using deep reinforcement learning, however, focused primarily on single agent revenue maximization and not solving for multi-agent Nash equilibria. \cite{Schlosser:2018} apply stochastic dynamic programming in a simulated environment to estimate consumer demand and maximize expected profit, however, do not consider the equilibrium solution of the market. \cite{vandergeer:2018-dpc} model market demand using parametric functions, with unknown parameters, yet they do not consider a Nash equilibrium nor the actions of previous pricing on demand. Thus, in almost all recent research on the dynamic pricing problem under uncertainty, there exists a desideratum to compute and promote Nash equilibria for market pricing strategies. Such strategies can be applied to eCommerce platforms such as Amazon auto pricing, allowing firms to opt for an automated pricing algorithm. 

A large body of literature on dynamic pricing has assumed an unknown demand function and tried to learn the intrinsic  relationship between price and demand. \cite{harrison2012bayesian} showed that a myopic Bayesian policy may lead to incomplete learning and poor profit performance. \cite{liu2020dynamic} studied a joint pricing and inventory problem while learning the price-sensitive demand using a Bayesian dynamic program. Many recent studies have revolved around non-parametric methods. \cite{besbes2009dynamic} developed a pricing strategy that balances the trade-off between exploration and exploitation. Furthermore, in \cite{Liu:2019-dyp} the demand function was a black box system where the demand was approximated from experience replay and live online testing in a reinforcement learning framework.

Dynamic pricing in oligopolies presents multiple problems. One aspect is the optimization problem, where we learn a Nash equilibrium policy when presented with the parameters of the market demand function, and second is the learning of the demand parameters itself. For example, \cite{den2014simultaneously} and \cite{keskin2014dynamic} studied parametric approaches, using maximum likelihood to estimate unknown parameters of the demand function. Recent research has been conducted in demand learning, for example in cases where multiple products exist, and firms can learn to price on an assortment of items \cite{ferreira:2022demand} using multinomial logit regression, or when multiple products exist in a demand learning scenario, and a limited inventory exists, Multi-Armed Bandit approaches have shown to be effective at devising profitable policies \cite{ferreira:2018thompson}. However, these approaches do not consider past pricing actions influencing the future market demand parameters. Furthermore, they do not consider competing firms aiming for a Nash equilibrium.

We propose a demand market where past pricing affects the future market demand in a market competition. To search for a Nash equilibrium, we apply a multi-agent Markov Decision Process, where the state transition is driven by the joint pricing actions of all players (or firms). To simplify the problem, we assume that each firm has unlimited inventory, and there is only a single product with no substitutes.

% establishing performance bounds. \cite{lei2014near} provided further improvement by removing the logarithmic terms in the upper bound.
% They proposed two modifications of the myopic Bayesian policy that avoid incomplete learning, and prove bounds on their performance.

% Intro: Markov Game Nash Section

\subsection{Multi-agent Markov Decision Processes}

Multi-agent Markov Decision Processes, or \textit{Markov Games} (MG's), constitute an important area of research, especially when multiple agents are self-interested, and an exact or approximate Nash equilibrium for the system is desired. The computation of Nash equilibria additionaly presents great difficulty when searching over the enormous joint state action space of the problem, and approximations to this search problem do exist \cite{Porter:2008}. Moreover, the existence of multiple Nash equilibria can further complicate the solution, as some solutions may be more desirable than others. 

Yet approximate search algorithms require prior knowledge of the joint reward function and are often limited to two players, modelled by a \textit{best response function}. We treat our payoff function as an \textit{oracle} \cite{Vorobeychik:2008}, meaning we have knowledge regarding the parametric form of the payoff function, however, there is no knowledge to the agents in the MG regarding how the reward function generates rewards for the players. The state of the game is visible to the agents, yet the agents have no knowledge regarding the state transition function. Nevertheless, the market demand parameters and Nash equilibria are known for purposes of experimentation. This eliminates the need for pre-period model fitting to pre-estimate market demand parameters \cite{ferreira2016:6analytics} allowing us to compare our solution to the theoretical solution, as opposed to an empirical estimate.

% The convergence behaviour of model free solvers in idealized systems to a Nash equilibrium, usually of a small scale and limited to two players have proven in the past to be convergent on the order of $O(1/\sqrt{T})$ \cite{Farina:2021} \cite{Kozuno:2021}. 

% two-time scale
Model-based approaches furthermore exist to provably find the existence of a Nash equilibrium \cite{Zhang:2020}. However, we aim for a Nash equilibrium solver which is model-free, where no knowledge regarding the reward or transition function is known to the agents. In recent literature, \cite{Raponi:2021} present a model-free algorithm to solve competitive MDP's when the environment parameters can be altered by an external party - but this is not always the case in market economics. Furthermore, \cite{Sayin:2021} propose a model-free decentralized deep learning framework, where the agent is blind to the other agents actions, and convergence to a Nash equilibrium is proven. \cite{Guan:2021} propose a convergent solution for zero-sum MG's via entropy-regularization. However, both \cite{Guan:2021} and \cite{Sayin:2021} impose a number of theoretical restrictions on the MG in order for this convergence to occur. Moreover, \cite{Kozuno:2021} present a provably convergent MG solver for imperfect information restricted to two agents. Nevertheless, we are concerned with an approximate solution to a full information MG for N agents. % In our environment, the actions of the other agents are known, however, the mechanics of the game are unknown, and the solver algorithm is less free of

% Intro: Deep learning

\textbf{Contributions:} This work outlines a methodology for Deep Q Learning, as introduced in \cite{mnih:2015}, by extending the framework to multi-agent reinforcement learning (MARL) with a Nash equilibrium objective based on the methods in \cite{hu:2003} and \cite{wang:2003-mgames}. Although the framework presented in \cite{hu:2003} is theoretically sound, the solution to the Nash equilibrium function is often intractable. We therefore apply approximation methods to compute a Nash equilibrium. Black box optimization is applied to estimate an $\epsilon$-Nash Equilibrium policy, and this approximation is learned as a series of neural networks. This MARL model is then applied to the domain of dynamic pricing.

\section{Dynamic Pricing Game} \label{sec:dynamic-pricing-game}

An oligopoly across $N$ firms is modelled as a multi-agent Markov Decision Process with a fully observable state action space. The state space is represented by the fully-observable demand-influencing reference price $\bar{p_t}$ being the state of the game $s^t$. In discrete intervals at time $t$, each agent issues a price for the item to be sold to the general market, $x^t_n$. The reference price, of the market at time $t$ is determined by a stochastic function of all the agents actions at time $t-1$. To focus the problem strictly on dynamic pricing, for each firm we assume a sufficiently large inventory, with no marginal costs of production, no possibility for inventory stockouts, no holding costs, and the capacity to always meet any realization of market demand.

% and immediate inventory order-up-to level at each time interval
% dictating the price driving the demand of the item at time $t$
\subsection{Markov Game Parameters} \label{sec:markov-game-params}

% , all values of $s_t$ where $t \geq 0$.
% , which can be simplified as $(s, x_1, . . . , x_N, r_1, . . . , r_N, s')$.
The joint action space $\{x_1^t \ ... \ x_n^t\} \in \mathbf{X}$ constitutes the current actions of all agents at time $t$, which drive the state transition by setting prices at time $t$ and represent the set price of the item from agent $n$ at time $t$. The state-action-reward space is defined as a tuple $(s^{t}, x_1^{t}, . . . , x_N^{t}, r_1^{t}, . . . , r_N^{t}, s^{t+1})$ where $r_n^{t}$ is the reward for agent $n$ at time $t$. The joint reward can be written as $\bar{r}_t = (r_1^{t}, . . . , r_N^{t})$, and the joint action can be written as $\bar{x}_t = (x_1^{t}, . . . , x_N^{t})$. $N$ denotes the number of agents. The exact transition probabilities are not known to the agents. Each agent must learn the demand function and optimal strategy as the MG progresses. A state $s^{t}$ is determined by the reference price of the market observable to all agents, $s \in \mathbb{R}$. We discretize the action space into even segments representing a firm's price. Where each segment represents an action $x_n \in X_n$, and $X_n$ is the action space for any agent $n \in N$. % Each agent $n \in N$ has an associated reward function $r_n: s\times x_1 \times ... \times x_N \to r_n$. State transition is controlled by the current state and one action from each agent: $s\times x_1 \times ... \times x_N \to s'$. Bounds on the state space and action space are limited to $|S| = |X_n| = 10$, that is there are 10 discrete intervals for each state $s^t$ and action $x_n^t$, representing price values.

\subsection{Reference Pricing} \label{sec:refernce-pricing}

% In an oligopoly market, the demand is influenced by the general market price among $N$ firms, referred to as the \textit{reference price}. 

% The perception about the value of a pre-existing market price, is one of the most important factors that influence customer decisions \cite{mazumdar2005reference}, numerous studies corroborating hypothesis are detailed \cite{fibich2003explicit}, \cite{popescu2007dynamic}, and \cite{heidhues2014regular}. 

% In order to demonstrate that a provable Nash equilibrium can be obtained by a machine learning algorithm, 
% In order to demonstrate that a provable Nash equilibrium can be obtained by a machine learning algorithm, we adopt a market model from \cite{taudes2012integrating},

The first pillar of any market dynamic is constituted by the \textit{demand function}, dictated by both the historical and contemporary prices, of the firms in the market. Demand processes can be ideally linear and continuous, however may not be guaranteed to be stationary or continuous \cite{vdenboer:2020-discon}, and can be subject to various exogenous factors such as environmental carryover effects \cite{rao1993pricing}. 

We create an idealization of a market based on \cite{taudes2012integrating}, where a dynamic pricing problem, with reference price effects and a linear demand function, is imposed. The expected current demand of a product is a function of the average price set by all firms, denoted as $\tilde{x}$, and a historical reference price, $\bar{p}$. The reference price $\bar{p}$ is given, and cannot be modified during the current time $t$, moreover, it is a function of the immediate past, $t-1$. Although \cite{taudes2012integrating} is not necessarily the only model that incorporates reference pricing \cite{mazumdar2005reference}, \cite{fibich2003explicit}, \cite{popescu2007dynamic}, \cite{heidhues2014regular}, we adopt it due to its simplicity and the existence of provable Nash equilibria in an oligopolistic framework.

% We create an idealization of a market system based on \cite{taudes2012integrating} for our simulation purposes.  

% when the theoretical optimum is unknown to the learning system.

\begin{align} % \label{eq:taudes-demand-noise}
    f(\tilde{x}) &= \beta_0 + \beta_1 \tilde{x} + \beta_2(\tilde{x} - \bar{p})\\
    \beta_0  &\geq 0, \ \beta_1 < 0, \ \beta_2 \leq 0, \ f(\tilde{x}) \geq  0, \tilde{x} \geq 0  \label{eq:taudes-demand}  % + \epsilon_D 
    % \mathbf{E}[\tilde{f}(\tilde{x})] &= \beta_0 + \beta_1 \tilde{x} + \beta_2(\tilde{x} - \bar{p})\\ 
    % \beta_0  &\geq 0, \ \beta_1 < 0, \ \beta_2 \leq 0, \ f(\tilde{x}) \geq  0, \tilde{x} \geq 0 \label{eq:constraints-fx-begin} 
\end{align}

$\epsilon_D$ is defined as the noise from a Poisson process. Such a Poisson process has the arrival rate $\lambda_t(\tilde{x}) = f(\tilde{x})$ from Eq. \eqref{eq:taudes-demand}, and standard deviation $\sigma = \sqrt{\lambda_t(\tilde{x})}$. Furthermore, \cite{taudes2012integrating} make the stipulation of decreasing demand with respect to increasing price, as illustrated in Condition \eqref{eq:taudes-demand}.

% Market demand is typically affected by a variety of factors, one of which is namely the reference price of a product, indicating the market perception of what the price of a product should be from the consumer point of view. 

This demand-influencing reference price can be affected by a variety of factors, such as inflation, unemployment, psychological perception \cite{raman:2002}. Moreover, in many proposed oligopoly models, as in \cite{janiszewski:1999range} and \cite{briesch:1997comparative}, the reference price is dictated by a historical joint market price of multiple firms. However, modelling a competitive market oligopoly with an autocorrelated reference price in a MG setting is not heavily investigated until now. In our model, we focus on designing a market whose reference price is driven by the historical $t-1$ joint market price, and additional factors which also affect the reference price are represented as noise $\epsilon_D$. Thus the transition of the reference price is determined by the average of the previous joint prices plus some Guassian randomness, $\epsilon_D$.

In our experiment, the reference price of a product is determined by the previous actions of the firms. We express the reference price function as $\bar{p}(t): \ \mathbb{R}^{N} \xrightarrow[]{} \mathbb{R}$, mapping a vector of historical pricing actions $\mathbf{x}^{t-1}$ to $\mathbb{R}$.  In the beginning of the Markovian process, the reference price is randomly generated within the action space of the pricing range. The stochastic nature of the market price transition entail the Markov property of the MG.

% The status quo function $\Omega(\mathbf{x^t})$ updates the reference price $\tilde{r}^{\ t}$ at time $t$. We propose an exponential smoothing function to model the behaviour of the reference price in Eq. \eqref{eq:ref-price-eq1}. This method takes the exponentially smoothed value of the previous price and the mean of the current firm prices. 

% \begin{equation} \label{eq:status-quo-function}
% \Omega(\mathbf{x_t}) \xrightarrow[]{} \tilde{x}_t,  \quad \forall \mathbf{x_t} \in \mathbb{R}^{N}
% \end{equation}

\begin{align} 
  % \Omega(\mathbf{x_t}) &= \alpha \Omega(\mathbf{x_{t-1}}) + (1-\alpha)\frac{1}{N} \sum_{\mathbf{x_t}} x_{t}^n + \epsilon_D \label{eq:ref-price-eq1} \\
  % \text{where} \quad \Omega(\mathbf{x_0}) &= \frac{1}{N} \sum_{\mathbf{x_0}} x_{0}^n, \quad 0 \leq \alpha < 1 \\
  \bar{p}(t) = \frac{1}{N} \sum^{N}_{n=1} x^{t-1}_n + \epsilon_D, \quad \text{where} \quad \bar{p}(t=0) = \frac{1}{N} \sum^{\mathbf{x^0}}_{n=1} x^{0}_n + \epsilon_D
\end{align} 

% Alternatively $\Omega(\mathbf{x_t})$ can also be a memoryless process, with the reference price being the minimum price at time $t$. 

% \begin{equation} \label{eq:ref-price-eq2}
% \Omega(\mathbf{x_t}) = \min(\mathbf{x_t}),  \quad \forall x_{t}^n \in \mathbf{x_t}
% \end{equation}

% We provide a demand function, $\lambda(\tilde{x}_t)$, derived from real eCommerce sales data (TBD-citation). 

% Depending on the formulation of $\Omega(\mathbf{x_t})$ the state transition of the market game can be driven by the pricing actions of each player, as well as the change of inventory after each completed transaction, where $\epsilon_D$ is some Gaussian noise over the otherwise deterministic function - creating a stochastic transition.

\subsection{Probabilistic Demand Function} \label{sec:profit-func}

The expected profit for player $n$ is $\mathbf{E}[\Pi_n(x_n)]$, which equals revenue as the marginal cost is of production is assumed to be 0, is defined as, 

\begin{align} % \label{eq:profit}
    \mathbf{E}[\Pi_n(x_n)] &= \Phi_n(x_n, x_{-n}) x_n f(\tilde{x}) \label{eq:profit-func1} \\
    \Phi_n(x_n, x_{-n}) &= \frac{ e ^{\phi(x_n)} }{\sum^{N}_{i=1} e^{\phi(x_i)}  } \label{eq:softmax-elas}
\end{align}

% provides a well-defined method to assign probabilities to arriving customers on whether they will make a purchase from a respective firm or not, 
$\Phi_n(x_n, x_{-n})$ in Eq. \eqref{eq:softmax-elas} represents the probability that a customer purchases the item sold by firm $n$, where the player $n$ prices their merchandise with price $x_n$. $f(x_n)$ is the expected demand during the single stage game for a fixed time period. Following the quantal response function from \cite{luce:1959} and \cite{goeree:2020}, we define a \textit{purchase elasticity function}, $\phi_n(x_n, x_{-n})$,

\begin{align} 
  \phi(x) = & 
    \begin{cases} 
      b_n - a_n x
        &\text{for \quad} 0 < x < b_n / a_n \\ 
      0 &\text{for \quad}  x > b_n / a_n 
    \end{cases} \label{eq:puchase-elas-function} \\
    \text{where} \quad &  x \in \mathbb{R}, x > 0, a \geq 0, b_n = 1 \label{eq:purchase_elas_price_constraints}
\end{align} 

In $\phi(x)$, $b_n$ represents the maximum weighted contribution to the probability of an item being purchased by a customer given a price. When the price $x$ is 0, this measure is $b_n$. For simplification, we assume the linear marginal decline of this measure of all players $a_n$ as equal, $a_n = a$, that is the market has the same elasticity regarding the marginal decrease of the customer willingness to purchase from any firm as price increases, with a negative slope $a$. The probability of a customer choosing to purchase from firm $n$ at price $x_n$ among prices set by other firms $x_{-n}$ is defined by combining a softmax function and purchase elasticity function, $\phi(x)$. This mechanism will prevent a solution where each firm undercuts each other to set their prices to 0, as the lowest price need not guarantee that a consumer will buy their product.

\section{Nash Equilibrium Conditions} \label{sec:eq-setting-main}

Computation of exact Nash, or approximate $\epsilon$-Nash, equilibria is an NP-hard problem and notoriously difficult to compute \cite{Daskalakis:2006}. This involves a search over the entire joint state action space, computing the value of each state under a candidate policy. Furthermore, it involves knowledge of the joint Q function and the transition probability of the system. In our scenario, we have the condition that all agents are identical, therefore the solution of one agent can apply to the solution of another. 

\subsection{Theoretical $\epsilon$-Nash Equilibria}

Under the market outlined in Section \ref{sec:dynamic-pricing-game}, given the market inputs $\bar{p}$ and $\tilde{x}$, in any pricing strategy, there exists an optimal deviation $d^* \in \mathbf{R}$, such that $\mathbf{E}[\Pi_n(x_n - d^*)]$ will yield the maximum profit advantage $\epsilon$.

% Suppose the market is in a situation where, an agent prices the item at $x_n$, and the market parameters are $\bar{p}$ and $\tilde{x}$. We show that there exists an optimal

% there exists deviation $d^* \in \mathbbf{R}$. If an agent $n$ a price of $x_n = \tilde{x} - d^*$ results in the optimal amount to undercut the current market price such that a maximum profit increase is gained by a margin of $\epsilon$.

\begin{align}
    \epsilon &= \underset{d^* \in \mathbf{R} }{\mathrm{max}} \ \Bigg(  \mathbf{E}[\Pi_n(x_n - d^*)] - \mathbf{E}[\Pi_n(x_n)] \Bigg) \label{eq:nash-def-diff} 
    % \epsilon &= \mathbf{E}[\Pi_n(x_n - d^*)] - \mathbf{E}[\Pi_n(x_n)]  \\
    % \epsilon &= \mathbf{E}[\Pi_n]^0 [\Omega(d^*) - 1]  \label{eq:eps-d-main} \\
    % \Omega(d) &= \frac{\mathbf{E}[\Pi_n]'}{\mathbf{E}[\Pi_n]^0}
    % \Omega (d) &= \Phi_d (1 + \frac{\gamma}{f(x_N)} d) (1 - \frac{1}{x_N} d)\label{eq:eps-range-d}
\end{align}

An increase in the individual profit of an agent $n$ from unilaterally deviating is denoted as $\epsilon$. Given this optimal deviation, $d^*$, a maximum theoretical upper bound on the profit advantage $\epsilon$ that can be obtained. Eq. \eqref{eq:d-star-main} provides the theoretical value of $d^*$ which is a function of both reference price $\bar{p}$ and market price $\tilde{x}$, as well as the fixed market parameters $\beta_0, \beta_1, \beta_2$. A derivation of $d^*$ can be found in Appendix \ref{sec:eq-setting}.

\begin{align}
    d^* &= \frac{ \sqrt{c_1^2 - c_1 + 4(c_2-1)c_2 - 2c_2} }{2c_2}  \label{eq:d-star-main} \\
    \text{where} \quad  c_1 &= \frac{- (\beta_1 + \beta_2)}{f(\tilde{x})} - \frac{1}{\tilde{x}}, \quad c_2 = \frac{- (\beta_1 + \beta_2)}{f(\tilde{x})\tilde{x}} \nonumber
\end{align}

\begin{figure}[!htb]
\minipage{0.4\textwidth}
  \includegraphics[width=\linewidth]{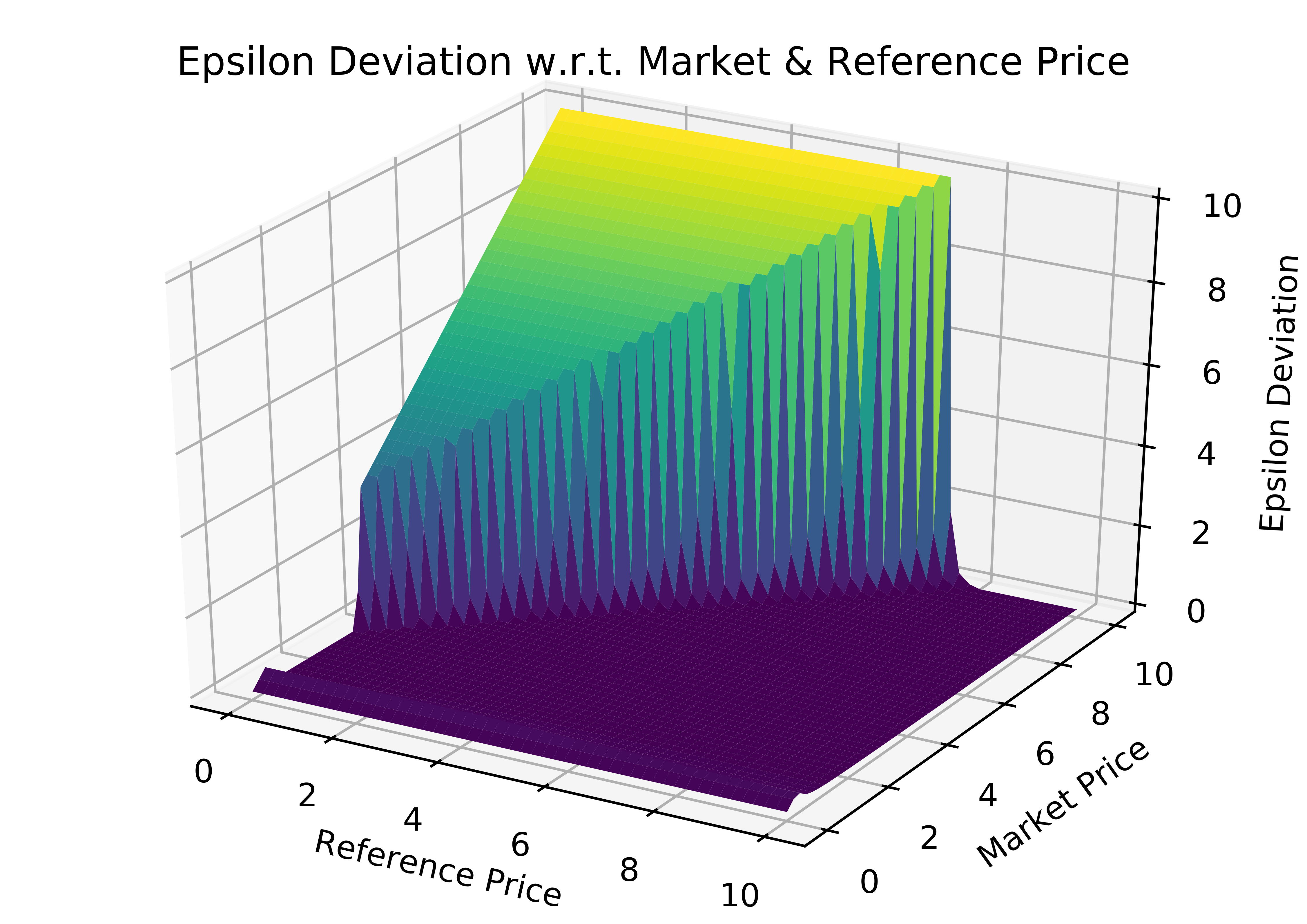}
  \caption*{\textbf{Market Scenario 1:} \\ $\beta_0 = 15, \beta_1 = -1.05, \beta_2 = -3.1, a = 0.1$}%\label{fig:market3}
\endminipage\hfill
\minipage{0.4\textwidth}
  \includegraphics[width=\linewidth]{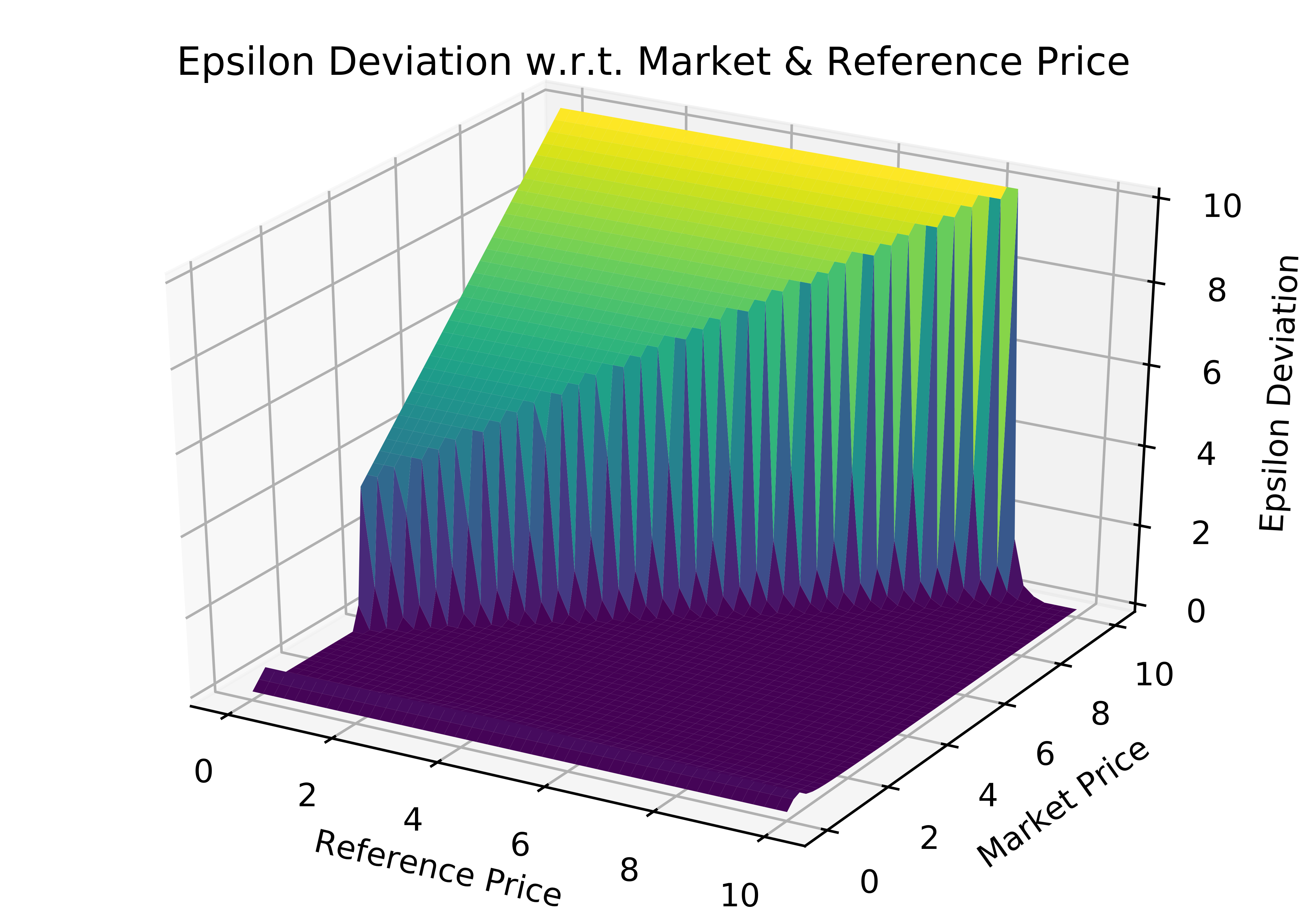}
  \caption*{\textbf{Market Scenario 2:} \\ $\beta_0 = 25, \beta_1 = -0.6, \beta_2 = -6.1, a = 0.1$}%\label{fig:market2}
\endminipage\hfill
\caption{Surface plot of potential advantage from deviation ($\epsilon$-deviation) from Nash equilibrium with respect to market price $\tilde{x}$, and reference price $\bar{p}$, with their respective market parameters $\beta_0, \beta_1, \beta_2, a$ for arrival of a single sales event. } \label{fig:market-topo-main}
\end{figure}

% Surface plot of $\epsilon$ deviation from Nash equilibrium with respect to market price $x_N$, and reference price $\bar{p}$ (with parameters

% \begin{figure}[!h] 
%     \begin{center}
%         \includegraphics[width=380]{market_nash_topology.png}
%           \caption{Surface plot of $\epsilon$ deviation from Nash equilibrium with respect to market price $x_N$, and reference price $\bar{p}$ (with market parameters $\beta_0 = 30, \beta_1 = -1.1, \beta_2 = -2, a = 0.1$).}
%           \label{fig:nash-topology}
%     \end{center}
% \end{figure} 

Using the Eq. \eqref{eq:d-star-main}, as visualized in Fig. \ref{fig:market-topo-main}, multiple Nash equilibria can exist when $\epsilon = 0$, or when $\epsilon$ is sufficiently small (see Section \ref{sec:results}). Such behaviour occurs when one agent unilaterally deviates from $\tilde{x}$, under reference price $\bar{p}$. Thus, the lower plateau from Fig. \ref{fig:market-topo-main} represents the region where a theoretical Nash Equilibrium exists. Bounds on the state space and action space are limited to $|S| = |X_n| = 10$, that is there are 10 discrete intervals for each state $s^t$ and action $x_n^t$, representing price values.

\subsection{Nash Equilibrium in Markov Games ($\delta$)}

The value of a policy, at each state, can be represented by a joint action which maximizes the joint Q function \cite{Watkins:1992} under policy $\pi(s)$. The probability of  agent $n$ taking action $x_n$ is defined as $\pi_n(s, x_n)$.

\begin{align} 
    v(s, \pi_n, \pi_{-n}) = \underset{\bar{x} \in \mathbf{X} }{\mathrm{max}} \ \prod_{n=1}^N \pi_n(s, x_n) Q_n(s, x_{n}, x_{-n}) \label{eq:val-joint-policy}
\end{align}

% The value of a policy can also be exactly determined if the transition and reward functions are known \cite{filar:1997}.

% \begin{equation} 
%     \mathbf{v}_\gamma (\pi_n, \pi_{-n}) = [\mathbf{I} - \gamma \mathbf{P}(\pi_n, \pi_{-n}) ]\mathbf{r}(\pi_n, \pi_{-n})  \label{eq:value-policy-linalg}
% \end{equation} 

An $\epsilon$-Nash equilibrium is defined as a joint policy such that the reward of a single stage game will not result in a greater payoff to the agent $n$ by more than $\epsilon$, when any agent unilaterally deviates from said joint policy. Provided $\epsilon$, we consider a bound on the corresponding MG, $\delta$, which defines a bound on the gain of the \textit{value} of a policy, $v(s, {\pi_{n}}^*, \pi^*_{-n})$, should agent $n$ unilaterally deviate with an alternative policy. The solution to Eq. \eqref{eq:val-joint-policy} can be computed by searching over the joint action space $\{ x_n, x_{-n} \}$.

\begin{align} 
    &\underset{\bar{x} \in \mathbf{X} }{\mathrm{argmax}} \prod_{n=1}^N \pi_n(s, x_n) Q_n(s, x_{n}, x_{-n}^*) \\
    &\leq \underset{\bar{x} \in \mathbf{X} }{\mathrm{argmax}} \prod_{n=1}^N \pi_n(s, x_n) Q_n(s, x_{n}^*, x_{-n}^*) + \delta \label{eq:nash-cond-q} \\
    v(s, \pi_n, \pi_{-n}^*) &\leq v(s, \pi_{n}^*, \pi_{-n}^*)  + \delta, \quad \forall s \in \mathbf{S} \label{eq:nash-cond-v} 
\end{align}

% The conditions presented Inequality \eqref{eq:nash-cond-q} and \eqref{eq:nash-cond-v} often presents an intractable problem for solving the Nash Equilbria conditions, we therefore opt for an approximation via gradient free optimization. 

$\delta$ serves as an upper bound on $\epsilon$, therefore the minimization of $\delta$ will also minimize any possible existence of $\epsilon$ for each single stage game in the MG. In fact, the existence of $\epsilon$ implies the existence of upper bound $\delta$, we provide a proof of this existence in Appendix \ref{sec:eps-delta-exis}.

% Furthermore, in Inequality \eqref{eq:nash-cond-q} and \eqref{eq:nash-cond-v}, $\delta$ represents the upper value bound on approximate Nash equilibrium on the value of a policy, whereas $\epsilon$ represents the upper bound on the value of a single stage game. 

\section{Multi-Agent Nash Q Learning} \label{sec:ma-ql}

% An agent takes on policy $\pi_n$, where the possible joint action $\bar{x}$ maximizes the Nash Q value times the joint probability of taking $\bar{x}$ for each state $s$. 

In our model, the game provides full information, where information regarding the state and the actions of all agents are visible to any agent. This allows for \textit{experience replay} for Q learning \cite{mnih:2015} to occur, and the Q function can be shared for all agents as they are assumed to be identical. The joint action space $\bar{x} = (x_1, ..., x_N)$ is defined as the combined actions of all agents at time $t$. Normally, in Q learning, the update mechanism searches for the action that maximizes the Q function, however, in \textit{Nash Q learning}, we must update the Q function with the solution to the Nash equilibrium. In our model, we represent the tabular Q function via a function approximator, which is denoted simply as $Q(s, \bar{x})$. As new experiences are obtained when the MG is played, the Nash Q value is updated. We utilize the Q update mechanism defined in \cite{hu:2003} as the update mechanism for the Nash Q estimator for Q learning. Given a Q function and a Nash policy, we search over the joint action that maximizes the scaled Q function \cite{laumonier:2005-multiagent}, as in Eq. \eqref{eq:nash-q-max}. In our representation, the Q function is a vector, returning the respective Q value for each agent, based on the joint probability input for the joint policy.

% \begin{align} \label{eq:joint-action-space}
%     \bar{a} = (a^1, ..., a^N)
% \end{align}

\begin{align} 
    Q'(s, \bar{x}) &\xleftarrow[]{} (1-\alpha) Q(s, \bar{x}) + \alpha [r + \gamma \mathcal{N}(s') ] \label{eq:q-update} \\
    \bar{x}^* &= \underset{\bar{x}}{\mathrm{argmax}} \ Q(s', \bar{x}) \prod_{i=1}^N \pi_n^*(s', x_n) \label{eq:nash-q-max}\\
    \mathcal{N}(s') &= Q(s', \bar{x}^*) \prod_{i=1}^N \pi_n^*(s', x^*_n)  \label{eq:nash-operator}
    % Q(s') &= \sum_{\bar{x} \in A} Q(s', \bar{x}) \label{eq:q-s}
\end{align}

% $\pi(s', \bar{x})$ represents the probability of taking joint action $\bar{x}$ while being in state $s'$ under joint policy $\pi$.
$\mathcal{N}(s')$ represents a \textit{Nash Operator}, indicating the Q value of a Nash equilibrium, this is approximated by a scaling factor, according to the Nash equilibrium policy, $\pi^*$, multiplied by the Q value of the joint action, $Q(s, \bar{x})$. Extending from the work of \cite{hu:2003}, we introduce a neural network for mapping any state action combination to its Nash policy, where the Nash scaling factor derives from and is used to compute $\mathcal{N}(s')$ in conjunction with the Q function. The Q-function used in Eq. \eqref{eq:nash-operator} and \eqref{eq:q-update} is represented by a deep neural network, simply referred to as the Nash Q Net $Q(s)$. In a full information game with homogenous agents, a shared neural network representing $Q(s, \bar{x})$ can be used for all agents in the Markov Game (if not homogeneous a separate Q network must be retained and updated separately for each agent). The Q Network parameters representing $\mathcal{N}(s')$ is learned in the same method as Deep Q learning \cite{mnih:2015}, the key innovation, is that the scaling factor to compute $\mathcal{N}(s')$ is obtained via a Nash Policy Net, $\Psi(s)$ (defined later in Section \ref{sec:nash-pol-learn}).

% In \cite{hu:2003} $\pi$ is a policy for joint Nash equilibrium $(\pi^*_n ... \pi^*_N)$.
% Efficient representation maps a state $s$ (input) to a collection of $Q(s, a)$ (outputs).
\subsection{Estimating Value Advantage via Black Box Optimization}

We apply a deep neural network to represent a mapping of a joint policy to its respective $\delta$ from Eq. \eqref{eq:nash-cond-v}, designated as $\Gamma(\pi) \mapsto \delta$, where $\delta$ is a vector containing the $\delta_s$ of each state. As the gradient of $\Gamma(\pi)$ cannot be easily evaluated, we apply gradient free black box optimization for $\delta$ minimization. Trust Region Optimization (TRO) has been shown to be effective for solving high-dimensional optimization problems \cite{wang:2021-lamcts} \cite{ericksson:2020-turbo} \cite{Diouane:2021-trego} \cite{Regis:2016}, particularly when the computational resources are available. To compute the existence of an $\epsilon$-Nash equilibrium in the high-dimensional policy space efficiently, we apply model-based Bayesian Optimization via Trust Region optimization (\texttt{TuRBO}) from \cite{ericksson:2020-turbo}. \texttt{TuRBO} combines standard TRO with a multi-armed bandit system via Thompson Sampling \cite{Thompson:1933} to optimize for local multiple trust regions simultaneously, with sufficient convergence time. However, the candidate generation step in \texttt{TuRBO} is not constrained to account for the valid joint probabilities of each agent, in which the sum of probabilities for each agent in its respective action space must sum to 1. We alter \texttt{TuRBO} by simply normalizing the original candidate probabilities over each set of probabilities belonging to an agent $n$. The resulting modified algorithm is denoted \texttt{TuRBO-p}, where any candidate value generated by \texttt{TuRBO-p} will have joint probabilities that sum to 1 for policies corresponding to each agent.

\begin{align} 
    V(s, \pi) &= \underset{\bar{x}}{\mathrm{max}} \ Q(s, \bar{x}) \prod_{i=1}^N \pi_n(s, x_n) \\
    \delta &= \underset{\pi_n'}{\mathrm{max}} \ \Bigg( V(s, \pi_n', \pi_{-n}) - V(s, \pi_{n}, \pi_{-n}) \Bigg) \quad \forall s \in \mathbf{S} \label{eq:delta-max}
    % \quad \forall s \in \mathbf{S}
\end{align}

The maximization problem is formulated in Eq. \eqref{eq:delta-max} representing the maximum gain in value of a policy, where agent $n$ deviates from policy $\pi_{n}$ with $\pi_n^{'}$. 

\subsection{Nash Policy Learning} \label{sec:nash-pol-learn}

The $\epsilon$-Nash policy is defined as a joint policy $(\pi^*_n, \pi^*_{-n})$ that minimizes $\delta$ from Eq. \eqref{eq:nash-cond-v}. Drawing inspiration from \cite{Ceppi:2010}, where a Nash equilibrium is found by effectively minimizing for $\epsilon$ via linear programming, we apply a similar technique of $\delta$-minimization. However, in \cite{Ceppi:2010}, the model parameters are known to the agents, and the Markov game was constrained to two players. In our MG, the joint reward function must be learned. Therefore, we perform approximate search over the policy space instead of using exact methods to discover any possible approximate Nash equilibrium policies. In principle, each state has maps to a corresponding probability in accordance with the Nash Policy which minimizes $\delta$ in Eq. \ref{eq:delta-max}. However, a table keeping track of such an enormous state policy space is not feasible. Therefore, we implement a Nash Policy Net $\Psi(s) \xrightarrow[]{} \pi^*$ to learn the state to policy mapping, which is the joint policy $\pi$ producing an approximate Nash equilibrium as approximated via \texttt{TuRBO-p}, designated as $\hat{\pi}^*(s)$.

\begin{align} 
    \hat{\pi}^*(s) = \underset{\pi}{\mathrm{argmin}} \ \Gamma(\pi)_s \label{eq:gamma-min}
\end{align}

\begin{algorithm}[h!]
\caption{Nash equilibrium learning}\label{algo:nash-eq-learning}
\begin{algorithmic}[1]

    \State Initialize state joint policy $\pi$.
    \State Initialize random parameters for $\Psi(s_t)$, $Q(s_t)$ and $\Gamma(\pi)$, as $\theta_{Q}$, $\theta_{\Psi}$, and $\theta_{\Gamma}$.
    \State Initialize MDP Environment, $\mathcal{M}$.
    
    \For {$e \in episodes $}:
        \State Get initial state $s_0$ from $\mathcal{M}$.
        \For {$t \in (0, t_{max})$}: Iterate until end of episode.
            % \State Joint Action $\bar{A} \xleftarrow[]{} \emptyset$
            % \quad \forall x_n \in \mathbf{X}
            \State Get action probabilities $\pi(s_t, x_n)$ from $\Psi(s_t)$.
                \For {$n \in (1, N)$}: Iterate through agents.
                    % \State $\{ {\pi_n^i, \pi_{-n}} \} \xleftarrow[]{} \mathtext{neighbour}(\pi_n^i) $
                    % \State Random sample action of each state policy. $a_n \thicksim \text{Multinomial} ( \pi_n^i(S) )$ 
                    \State Sample $x_n \thicksim Multinomial(s_t, \pi(\overline{x}))$ 
                \EndFor
            \State Obtain joint action $\overline{x} \xleftarrow[]{}  \{ x_1, ... , x_N \}$
            \State Assign  $\delta_s$ according to Eq. \eqref{eq:delta-max} via \texttt{TuRBO-p}.
            \State Find $\pi^* \xleftarrow[]{}  \underset{\pi}{\mathrm{argmin}} \ \Gamma(\pi, s_t)$ via \texttt{TuRBO-p}.
            \State Execute joint action $\overline{x}$ in $(\mathcal{M}, \ s_{t})$ to obtain $s_{t+1}, \overline{r}_t$.
            \State Append $(s_t, \overline{x}_t, \overline{r}_t, \epsilon_s, \pi, s_{t+1} )$ to experience replay $\mathcal{D}$.
            % \State Initialize $\mathbf{\hat{v}}_c \xleftarrow[]{} \Xi_m(\pi_n^i, \pi_{-n})$
            \State Update State $s_{t} \xleftarrow[]{} s_{t+1}$.

            \If {$|\mathcal{D}| > batchsize$}:
                \State Sample minibatch $d$ from $\mathcal{D}$.
                \State Set $Q'(s_t, \bar{x}) \xleftarrow[]{}  
                    \begin{cases} 
                      r_t &\text{for \ } s_t \ \text{terminal} \\ 
                      \text{Eq. \eqref{eq:q-update}} &\text{for \ }  s_t \ \text{not terminal} 
                    \end{cases} $
                
                    \State $\mathcal{L}_{Q} \xleftarrow[]{} (Q'(S_t, \bar{x}) - Q(S_t, \bar{x}))^2$ Nash Q Update.
                \State $\mathcal{L}_{\Psi} \xleftarrow[]{} (\hat{\pi} - \pi^*)^2$, where $\hat{\pi}$ is from $\Psi(s)$.
                \State $\mathcal{L}_{\Gamma} \xleftarrow[]{} (\hat{\epsilon} - \epsilon)^2$ Nash $\epsilon$ update, where $\hat{\epsilon}$ is from $\Gamma(\pi)$
                \State Backpropagate $\mathcal{L}_{Q}$, $\mathcal{L}_{\Psi}$ and $\mathcal{L}_{\Gamma}$ on $\mathcal{Q}$, $\Psi$, and $\Gamma$.
                \State Update $\theta_{Q}$, $\theta_{\Psi}$, and $\theta_{\Gamma}$ via gradient descent.
                
                % \State Epsilon learning
            \EndIf
            
            \State Update joint policy $\overline{\pi} \xleftarrow[]{} \overline{\pi}^*$.
        \EndFor
    \EndFor

\end{algorithmic}
\end{algorithm}

\section{Results} \label{sec:results}
% corresponding to the learning of the Nash Q function via deep reinforcement learning

% \begin{figure}[!htb]
% \minipage{0.4\textwidth}
%   \includegraphics[width=\linewidth]{plots/sc1_loss.png}
%   \caption*{Loss behaviour of Scenario 1 with $n=3$.}% \label{fig:market1_loss}
% \endminipage\hfill
% \minipage{0.4\textwidth}
%   \includegraphics[width=\linewidth]{plots/market2_loss_5n.png}
%   \caption*{Loss behaviour of Scenario 2 with $n=5$.}% \label{fig:market2_loss}
% \endminipage\hfill
% \caption{Decreasing loss behaviour of batch update during Deep Q Learning $Q$ (left y axis) and Nash Net update $\Psi(s)$ (right y axis) demonstrating that $\Psi(s)$ has learned a representation of the Nash policy (x-axis represents iterations multiplied by episode).} \label{fig:loss_plot_main}
% \end{figure}

\begin{figure}[!htb]
\minipage{0.4\textwidth}
  \includegraphics[width=\linewidth]{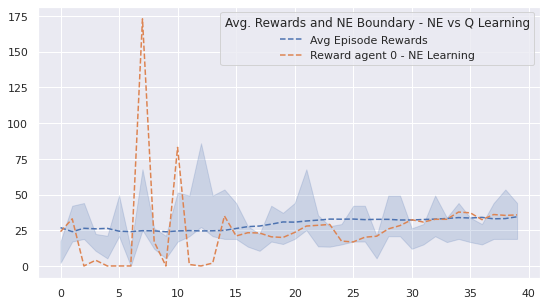}
  \caption*{Market rewards per episode Scenario 1, $n=3$.}\label{fig:market3_reward}
\endminipage\hfill
\minipage{0.4\textwidth}
  \includegraphics[width=\linewidth]{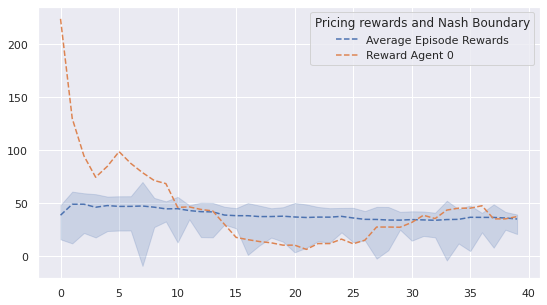}
  \caption*{Market rewards per episode Scenario 2, $n=3$.}\label{fig:market2_reward}
\endminipage\hfill
\caption{In a Nash equilibrium, both the market average reward (blue), and single agent reward (orange) should fall within the Nash equilibria boundary (blue shade), as the MG progresses. }\label{fig:reward_plots_main}
\end{figure}

\textbf{Stabilization of realized market rewards to Nash equilibrium:} Fig. \ref{fig:reward_plots_main} represents the convergence of the average market reward of a single agent (randomly selected as agent 0) towards a Nash equilibrium. We superimpose the boundaries of the theoretical Nash equilibrium reward over the plot. The reward is obtained from the boundaries of theoretical Nash equilibria, where $\epsilon < \epsilon_l$ ($\epsilon_l = 0.0001$). The topology of this function is illustrated in Fig. \ref{fig:market-topo-main}, and for each state, or reference price, there exists a boundary where the policy deviation of any agent can occur without significant unilateral reward gain. The expected reward is computed using the market model parameters per Scenario ($\beta_0, \beta_1, \beta_2, a$). From the expected reward, we compare it with the positive difference of theoretical equilibrium reward $\Pi' = f(x_n') \ x_n'$ where $x_n'$. This exists as two boundary points indicating the area where $\epsilon < \epsilon_l$. For each episode the average reward per timestep over the episode lengths is recorded. From Fig. \ref{fig:reward_plots_main}, we observe the average reward per agent of the system per episode (blue dashed line), and the average reward of a single agent per episode (orange dashed line), to converge within the boundary of Nash equilibrium (blue shade). However, due to randomization from black box optimization and/or stochastic behaviour of the Markov Game this expected reward can sometimes fall outside of the NE region.

We demonstrate empirically in Appendix \ref{sec:plots-figs}, that the loss function of $Q(s)$ and $\Psi(s)$ decrease with each RL episode. This merely indicates that a representation of the function mapping state to its corresponding Nash policy based on policy-to-$\epsilon$ model $\Gamma(\pi)$ is being learned, and a stable solution is proposed. Hyperparameters are presented in Appendix \ref{sec:hardware-hyperparam}.

\begin{figure}[!htb]
\minipage{0.4\textwidth}
  \includegraphics[width=\linewidth]{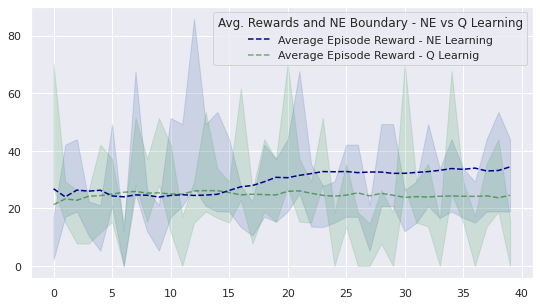}
  \caption*{Scenario 1 - NE Learning vs Q learning, $n=3$.}\label{fig:market3_max_gain}
\endminipage\hfill
\minipage{0.4\textwidth}
  \includegraphics[width=\linewidth]{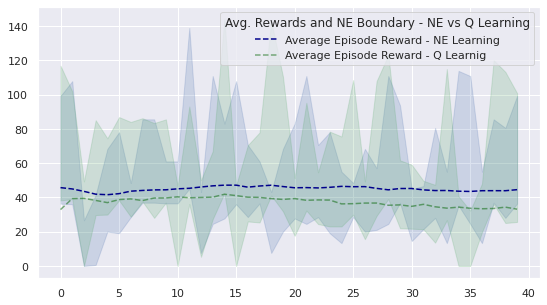}
  \caption*{Scenario 2 - NE Learning vs Q learning, $n=5$.}\label{fig:market2_reward}
\endminipage\hfill
\caption{Average revenue per episode from a NE learning model (blue) versus a baseline Q learning model (green), with NE reward boundaries overlayed in their respective colours.  }\label{fig:baseline_comp}
\end{figure}

\textbf{Comparison with baseline cooperative Q-learning model:} We compare our NE learning model wit a naive cooperative Q learning model to serve as a baseline measure of model performance. In the naive Q learning model, the objective is simply to maximize each agent's respective rewards, and avoid any kind of NE computation. The Q update function in Eq. \eqref{eq:q-update} is simply a maximization of the joint Q function, instead of using $\mathcal{N}(s')$. From Fig.\ref{fig:baseline_comp}, we generally observe lower average reward values from the baseline model compared to the NE learning model. The NE reward ranges remain very similar. From an economic perspective, the Nash Q learning agents take into account that other agents will compete against them, and the solver optimizes accordingly, whereas, in cooperative Q learning no consideration for the other agents' strategies are taken, and agents may to over-compete, driving average rewards lower in an oligopoly.

% $\Pi_\epsilon$ is the expected maximum reward any agent can obtain by deviating from joint policy of the other agents in the each state. From an economic perspective the learned policy of the NE learning provides less incentive for agents to undercut their competitors. And it follows, that when a policy is learned cooperatively without considering NE, the incentive to undercut your competitors is greater. Secondly, the NE MARL algorithm produces a larger NE boundary compared to the coop. Q learning algorithm. This indicates that the NE-MARL model learns a joint policy where more points of NE can exist providing a larger range of realized profit and still be in NE. Intuitively, under the learned joint policy, any individual agent needs to undercut the market significantly in order to achieve any significant gain in revenue. 

\section{Conclusion} \label{sec:conclusion}

We created a Markov Game that represents an n-firm oligopoly, based on previously established market models where theoretical $\epsilon$ bounds for a Nash equilibrium policy exist. A black box algorithm is applied to estimate an upper bound on the $\epsilon$ value of a joint policy, represented by neural network $\Gamma(\pi)$. Similarly a Nash Policy Net $\Psi(s)$ is learned to represent the $\epsilon$-minimizing policy from black box optimization, constituting an $\epsilon$-Nash policy. Thus $\Psi(s)$ can be used together with traditional Deep Q learning, to solve for a Nash equilibrium. Empirically, we show that the average reward of all agents, and the reward of a respective single agent, converges to an approximate Nash equilibrium. The limitations of this research are the limited action space of the agents, as well as the identical nature of the agents. Larger scale experimentation under this framework is suggested. Furthermore, the proposed market model could be enhanced by implementing more complex non-linear market oligopoly environments.

%%%%%%%%%%%%%%%%%%%%%%%%%%%%%%%%%%%%%%%%%%% HHH%%%%%%%%%%%%%%%%%%%%%%%%55

% \ack We would like to thank the referees for their comments, which helped improve this paper considerably

\bibliography{main}

\begin{thebibliography}{10}

\bibitem{bertrand:1883}
Joseph Bertrand, {\em Review of Walras's Théorie mathématique de la richesse
  sociale and Cournot's Recherches sur les principes mathématiques de la
  théorie des richesses},  73–81, Cambridge University Press, 1889.

\bibitem{besbes2009dynamic}
Omar Besbes and Assaf Zeevi, `Dynamic pricing without knowing the demand
  function: Risk bounds and near-optimal algorithms', {\em Operations
  Research}, {\bf 57}(6),  1407--1420, (2009).

\bibitem{briesch:1997comparative}
Richard~A Briesch, Lakshman Krishnamurthi, Tridib Mazumdar, and Sevilimedu~P
  Raj, `A comparative analysis of reference price models', {\em Journal of
  Consumer Research}, {\bf 24}(2),  202--214, (1997).

\bibitem{Ceppi:2010}
Sofia Ceppi, Nicola Gatti, Giorgio Patrini, and Marco Rocco, `Local search
  methods for finding a nash equilibrium in two-player games', pp. 335--342,
  (08 2010).

\bibitem{Daskalakis:2006}
Constantinos Daskalakis, Paul~W. Goldberg, and Christos~H. Papadimitriou, `The
  complexity of computing a nash equilibrium', in {\em Proceedings of the
  Thirty-Eighth Annual ACM Symposium on Theory of Computing}, STOC '06, p.
  71–78, New York, NY, USA, (2006). Association for Computing Machinery.

\bibitem{vdenboer:2020-discon}
Arnoud~V. Den~Boer and N.~Bora Keskin, `Discontinuous demand functions:
  Estimation and pricing', {\em Management Science}, {\bf 66}, (04 2020).

\bibitem{den2014simultaneously}
Arnoud~V den Boer and Bert Zwart, `Simultaneously learning and optimizing using
  controlled variance pricing', {\em Management Science}, {\bf 60}(3),
  770--783, (2014).

\bibitem{Diouane:2021-trego}
Youssef Diouane, Victor Picheny, Rodolphe~Le Riche, and Alexandre~Scotto
  Di~Perrotolo.
\newblock Trego: a trust-region framework for efficient global optimization,
  2021.

\bibitem{ericksson:2020-turbo}
David Eriksson, Michael Pearce, Jacob~R. Gardner, Ryan Turner, and Matthias
  Poloczek, `Scalable global optimization via local bayesian optimization',
  {\em CoRR}, {\bf abs/1910.01739}, (2019).

\bibitem{ferreira2016:6analytics}
Kris~Johnson Ferreira, Bin Hong~Alex Lee, and David Simchi-Levi, `Analytics for
  an online retailer: Demand forecasting and price optimization', {\em
  Manufacturing \& Service Operations Management}, {\bf 18}(1),  69--88,
  (2016).

\bibitem{ferreira:2022demand}
Kris~Johnson Ferreira and Emily Mower, `Demand learning and pricing for varying
  assortments', {\em Manufacturing \& Service Operations Management}, (2022).

\bibitem{ferreira:2018thompson}
Kris~Johnson Ferreira, David Simchi-Levi, and He~Wang, `Online network revenue
  management using thompson sampling', {\em Operations Research}, {\bf 66}(6),
  1586--1602, (2018).

\bibitem{fibich2003explicit}
Gadi Fibich, Arieh Gavious, and Oded Lowengart, `Explicit solutions of
  optimization models and differential games with nonsmooth (asymmetric)
  reference-price effects', {\em Operations Research}, {\bf 51}(5),  721--734,
  (2003).

\bibitem{goeree:2020}
Jacob~K. Goeree, Charles~A. Holt, and Thomas~R. Palfrey, {\em Stochastic game
  theory for social science: a primer on quantal response equilibrium}, Edward
  Elgar Publishing, Cheltenham, UK, 2020.

\bibitem{harrison2012bayesian}
J~Michael Harrison, N~Bora Keskin, and Assaf Zeevi, `Bayesian dynamic pricing
  policies: Learning and earning under a binary prior distribution', {\em
  Management Science}, {\bf 58}(3),  570--586, (2012).

\bibitem{heidhues2014regular}
Paul Heidhues and Botond K{\H{o}}szegi, `Regular prices and sales', {\em
  Theoretical Economics}, {\bf 9}(1),  217--251, (2014).

\bibitem{hu:2003}
Junling Hu and Michael~P. Wellman, `Nash q-learning for general-sum stochastic
  games', {\em Journal of Machine Learning Research}, {\bf 4},  1039–1069,
  (December 2003).

\bibitem{janiszewski:1999range}
Chris Janiszewski and Donald~R Lichtenstein, `A range theory account of price
  perception', {\em Journal of Consumer Research}, {\bf 25}(4),  353--368,
  (1999).

\bibitem{keskin2014dynamic}
N~Bora Keskin and Assaf Zeevi, `Dynamic pricing with an unknown demand model:
  Asymptotically optimal semi-myopic policies', {\em Operations Research}, {\bf
  62}(5),  1142--1167, (2014).

\bibitem{Kozuno:2021}
Tadashi Kozuno, Pierre Ménard, Rémi Munos, and Michal Valko.
\newblock Model-free learning for two-player zero-sum partially observable
  markov games with perfect recall, 2021.

\bibitem{laumonier:2005-multiagent}
Julien Laum{\^o}nier and Brahim Chaib-draa, `Multiagent q-learning: Preliminary
  study on dominance between the nash and stackelberg equilibriums', in {\em
  Proceedings of AAAI-2005 Workshop on Multiagent Learning, Pittsburgh, USA},
  (2005).

\bibitem{Liu:2019-dyp}
Jiaxi Liu, Yidong Zhang, Xiaoqing Wang, Yuming Deng, and Xingyu Wu, `Dynamic
  pricing on e-commerce platform with deep reinforcement learning', {\em CoRR},
  {\bf abs/1912.02572}, (2019).

\bibitem{liu2020dynamic}
Jue Liu, Zhan Pang, and Linggang Qi, `Dynamic pricing and inventory management
  with demand learning: A bayesian approach', {\em Computers \& Operations
  Research}, {\bf 124},  105078, (2020).

\bibitem{luce:1959}
R.~Duncan Luce, {\em Individual Choice Behavior: A Theoretical Analysis},
  Wiley, New York, NY, USA, 1959.

\bibitem{mazumdar2005reference}
Tridib Mazumdar, Sevilimedu~P Raj, and Indrajit Sinha, `Reference price
  research: Review and propositions', {\em Journal of Marketing}, {\bf 69}(4),
  84--102, (2005).

\bibitem{mnih:2015}
Volodymyr Mnih, Koray Kavukcuoglu, David Silver, Andrei~A. Rusu, Joel Veness,
  Marc~G. Bellemare, Alex Graves, Martin Riedmiller, Andreas~K. Fidjeland,
  Georg Ostrovski, Stig Petersen, Charles Beattie, Amir Sadik, Ioannis
  Antonoglou, Helen King, Dharshan Kumaran, Daan Wierstra, Shane Legg, and
  Demis Hassabis, `Human-level control through deep reinforcement learning',
  {\em Nature}, {\bf 518}(7540),  529--533, (February 2015).

\bibitem{Nash:1950}
John~F. Nash, `Equilibrium points in n-person games', {\em Proceedings of the
  National Academy of Sciences}, {\bf 36}(1),  48--49, (1950).

\bibitem{popescu2007dynamic}
Ioana Popescu and Yaozhong Wu, `Dynamic pricing strategies with reference
  effects', {\em Operations Research}, {\bf 55}(3),  413--429, (2007).

\bibitem{Porter:2008}
Ryan Porter, Eugene Nudelman, and Yoav Shoham, `Simple search methods for
  finding a nash equilibrium', {\em Games and Economic Behavior}, {\bf 63}(2),
  642--662, (2008).
\newblock Second World Congress of the Game Theory Society.

\bibitem{raman:2002}
Kalyan Raman, Frank~M Bass, et~al., `A general test of reference price theory
  in the presence of threshold effects', {\em Tijdschrift voor Economie en
  management}, {\bf 47}(2),  205--226, (2002).

\bibitem{Raponi:2021}
Giorgia Ramponi, Alberto~Maria Metelli, Alessandro Concetti, and Marcello
  Restelli, `Learning in non-cooperative configurable markov decision
  processes', in {\em Advances in Neural Information Processing Systems}, eds.,
  A.~Beygelzimer, Y.~Dauphin, P.~Liang, and J.~Wortman Vaughan, (2021).

\bibitem{rao1993pricing}
Vithala~R Rao, `Pricing models in marketing', {\em Handbooks in Operations
  Research and Management Science}, {\bf 5},  517--552, (1993).

\bibitem{Regis:2016}
Rommel~G. Regis, `Trust regions in kriging-based optimization with expected
  improvement', {\em Engineering Optimization}, {\bf 48}(6),  1037--1059,
  (2016).

\bibitem{Sayin:2021}
Muhammed~O. Sayin, Kaiqing Zhang, David~S. Leslie, Tamer Basar, and Asuman~E.
  Ozdaglar, `Decentralized q-learning in zero-sum markov games', {\em CoRR},
  {\bf abs/2106.02748}, (2021).

\bibitem{Schlosser:2018}
Rainer Schlosser and Martin Boissier, `Dynamic pricing under competition on
  online marketplaces: A data-driven approach', in {\em Proceedings of the 24th
  ACM SIGKDD International Conference on Knowledge Discovery and Data Mining},
  KDD '18, p. 705–714, New York, NY, USA, (2018). Association for Computing
  Machinery.

\bibitem{taudes2012integrating}
Alfred Taudes and Christian Rudloff, `Integrating inventory control and a price
  change in the presence of reference price effects: a two-period model', {\em
  Mathematical Methods of Operations Research}, {\bf 75}(1),  29--65, (2012).

\bibitem{Thompson:1933}
William~R Thompson, `{On the likelihood that one unknown probability exceeds
  another in view of evidence of two samples}', {\em Biometrika}, {\bf
  25}(3-4),  285--294, (12 1933).

\bibitem{vandergeer:2018-dpc}
Ruben van~de Geer, Arnoud~V. den Boer, Christopher Bayliss, Christine S.~M.
  Currie, Andria Ellina, Malte Esders, Alwin Haensel, Xiao Lei, Kyle D.~S.
  Maclean, Antonio Martinez-Sykora, and et~al., `Dynamic pricing and learning
  with competition: insights from the dynamic pricing challenge at the 2017
  informs rm \& pricing conference', {\em Journal of Revenue and Pricing
  Management}, {\bf 18}(3),  185–203, (Oct 2018).

\bibitem{Vorobeychik:2008}
Yevgeniy Vorobeychik and Michael Wellman, `Stochastic search methods for nash
  equilibrium approximation in simulation-based games', pp. 1055--1062, (01
  2008).

\bibitem{wang:2021-lamcts}
Linnan Wang, Rodrigo Fonseca, and Yuandong Tian, `Learning search space
  partition for black-box optimization using monte carlo tree search', {\em
  CoRR}, {\bf abs/2007.00708}, (2020).

\bibitem{wang:2003-mgames}
Xiaofeng Wang and Tuomas Sandholm, `Reinforcement learning to play an optimal
  nash equilibrium in team markov games', in {\em Advances in Neural
  Information Processing Systems}, eds., S.~Becker, S.~Thrun, and K.~Obermayer,
  volume~15. MIT Press, (2003).

\bibitem{Watkins:1992}
Christopher J. C.~H. Watkins and Peter Dayan, `Q-learning', {\em Machine
  Learning}, {\bf 8}(3),  279--292, (May 1992).

\bibitem{Zhang:2020}
Kaiqing Zhang, Sham~M. Kakade, Tamer Basar, and Lin~F. Yang, `Model-based
  multi-agent {RL} in zero-sum markov games with near-optimal sample
  complexity', {\em CoRR}, {\bf abs/2007.07461}, (2020).

\bibitem{Guan:2021}
Qifan Zhang, Yue Guan, and Panagiotis Tsiotras, `Learning nash equilibria in
  zero-sum stochastic games via entropy-regularized policy approximation', {\em
  CoRR}, {\bf abs/2009.00162}, (2020).

\end{thebibliography}

\clearpage
\section*{Supplemental Material}
\appendix

% \section{Appendix}
\section{Hardware Configurations and Model Hyperparameters} \label{sec:hardware-hyperparam}

All experiments were run on a computer with an Intel Core i5-10600K CPU processor at 4.10GHz, 32 GB DDR3 RAM, with an NVIDIA GeForce GTX 1080 8GB GDDR5X graphics card. The hyperparameters used for all experiments are listed in Table \ref{table:hyperparam}.

\begin{table}

\centering
\begin{tabular}{||c c||}
 \hline
 Parameter & Value \\ [0.5ex] 
 \hline\hline
 No. Hidden Layers $\Gamma(\pi)$ & 3 \\
  No. Hidden Layers $\Psi(s)$ & 3 \\
  No. Hidden Layers $Q(s)$ & 3 \\
Hidden Layer Size $\Gamma(\pi)$ & 1500 \\
Hidden Layer Size $\Psi(s)$ & 1500 \\
Hidden Layer Size $Q(s)$ & 75 \\
 No. Episodes & 40 \\
 Batch Update Frequency & 20 \\ 
 Batch Size & 10 \\ 
 RL Exploration Constant & 0.05 \\ 
Max. Steps per Episode & 30  \\ 
Discount Factor ($\gamma$) & 0.9\\ 
Learning Rate ($\alpha$) & 0.001\\ 
MDP Action Dimensions & 10 \\ 
\texttt{TuRBO-p} Max. Evaluations & 10 \\ 
\texttt{TuRBO-p} Batch Size & 4 \\[1ex] 
 \hline
\end{tabular}
\caption{Hyperparameters used for deep reinforcement learning. }
\label{table:hyperparam}
\end{table}

\section{Softmax Win Probability} \label{sec:win-prob}

\textbf{Proposition:} In an N-player game when deviating from the equilibrium market price by an amount of $d$, given the softmax win probability $\Phi(x_n)$ in Eq. \eqref{eq:softmax-elas}, the probability of winning a customer changes by a factor of $\Phi_d$ as defined in Eq. \eqref{eq:phi-def}.

\textbf{Proof:} Suppose N-1 players set an equal price of $\tilde{x}$ and one player deviates with price $x_n$, where $x_n = \tilde{x} - d$. 

The players who do not deviate from the equilibrium price, will have a win probability of $\Phi(\tilde{x})$

\begin{align} 
    \Phi(\tilde{x}) &= \frac{ e ^{1 - a\tilde{x}} }{\sum^{N}_{i=1} e^{1 - ax_i}} \label{eq:softmax-defn}
\end{align}

The player that deviates from the market price will have a win probability of  $\Phi(x_n)$,

\begin{align} 
    \Phi(x_n) &= \frac{ e ^{1 - ax_n} }{\sum^{N}_{i=1} e^{1 - ax_i}  } 
    = \frac{ e^{1 - a(\tilde{x} - d) } }{\sum^{N}_{i=1} e^{1 - ax_i} - e^{\tilde{x}} +  e^{1 - a(\tilde{x} - d)} } \nonumber \\
    &= \frac{ e^{1 - a(\tilde{x} - d) } }{N_{\Phi}  + e^{1 - a(\tilde{x} - d)} }, \quad \text{where} \ N_{\Phi} = \sum^{N}_{i=1} e^{1 - ax_i} - e^{\tilde{x}} \label{eq:phi-def}
\end{align}

The increase in win probability by deviating from the equilibrium price $\tilde{x}$ by an increment of $d$ is denoted by $\Phi_d$,

\begin{align} 
    \Phi_d &= \frac{\Phi(x_n)}{\Phi(\tilde{x})} 
    = \frac{ e^{1 - a(\tilde{x} - d) } } {N_{\Phi} + e^{\tilde{x} - d}} \frac{N_{\Phi} + e^{\tilde{x}}}{ e ^{1 - a\tilde{x}}   } = e^{ad} \frac{N_{\Phi} + e^{\tilde{x}}}{N_{\Phi} + e^{\tilde{x} - d}} \label{eq:softmax-gain}
    %\Phi_d &= \frac{\Phi(x_n)}{\Phi(\tilde{x})} = \frac{ e ^{1 + a\tilde{x} - d } }{\sum^{N}_{i=1} e^{1 - ax_i}  } \frac{\sum^{N}_{i=1} e^{1 - ax_i} }{ e ^{1 + a\tilde{x}}   } \label{eq:softmax-n-player-d} \\
\end{align}

As follows from Eq. \eqref{eq:softmax-gain} the probability of winning a customer's purchase by changing price with deviation $d$ with respect to the equilibrium price effectively changes the win probability by a factor $\Phi_d$.

\subsection{Admissible Values of Profit Function}

\textbf{Proposition:} In an N-player game when deviating from the market price by an amount of $d \in \mathbb{R}$, there always exists a boundary $(d^-, d^+)$ such that the expected profit from deviating $\mathbf{E}[\Pi_n]' > 0$ exists. We define this as the admissible range.

% is always greater than not undercutting $\mathbf{E}[\Pi_n]^0$, that is $\mathbf{E}[\Pi_n]^- > \mathbf{E}[\Pi_n]^0$. 

\begin{align} 
    \frac{\mathbf{E}[\Pi_n]'}{\mathbf{E}[\Pi_n]^0} > 0, \quad \forall d \in (d^-, d^+) \label{eq:eps-range-ded}
\end{align}

% $\Omega(d)$ represents the ratio of profit gain from undercutting $\mathbf{E}[\Pi_n]^'$ versus obeying the market price $\mathbf{E}[\Pi_n]^0$. 

\textbf{Proof:} We define the \textit{gain function} $\Omega(d)$ as, 

\begin{align} 
    \Omega(d) &= \frac{\mathbf{E}[\Pi_n]'}{\mathbf{E}[\Pi_n]^0} 
    = \frac{\Phi(x_n) f(x_n) x_n  }{\Phi(\tilde{x}) f(\tilde{x}) \tilde{x}} 
    = \frac{\Phi(\tilde{x})\Phi_d [f(\tilde{x}) + \gamma d] (\tilde{x} - d ) }{\Phi(\tilde{x}) f(\tilde{x}) \tilde{x}} \nonumber \\
    &= \Phi_d \frac{[f(\tilde{x}) + \gamma d] (\tilde{x} - d ) }{ f(\tilde{x}) \tilde{x}}
    = \Phi_d (1 + \frac{\gamma}{f(\tilde{x})} d) (1 - \frac{1}{\tilde{x}} d) \nonumber \\
    \Omega (d) &= \Phi_d (1 + \frac{\gamma}{f(\tilde{x})} d) (1 - \frac{1}{\tilde{x}} d)\label{eq:eps-range-d}
\end{align}

Given the deviation condition $\Omega(d) > 0$ and Condition \eqref{eq:taudes-demand}, the admissible range is more precisely defined as a range for $d$ where a solution for Inequality \eqref{eq:d-bound} exists,

\begin{align} 
    \Omega(d) \implies (1 + \frac{\gamma}{f(\tilde{x})} d) (1 - \frac{1}{\tilde{x}} d) > 0  &\implies -\frac{f(\tilde{x})}{\gamma} < d < \tilde{x} \label{eq:d-bound}
    % \mathbf{E}[\Pi_n]^- > \mathbf{E}[\Pi_n]^0 &\implies \text{where}  \label{eq:bound-on-d}
    %1 - \frac{1}{\tilde{x}} d &> &\implies 0 d < \tilde{x}
    %\text{where} \quad d &\in [0, \tilde{x}] \implies \mathbf{E}[\Pi_n]^- > \mathbf{E}[\Pi_n]^0
\end{align}

We see from Inequality \eqref{eq:d-bound} that there is a bound on admissible values of $d \in (-f(\tilde{x})/\gamma, \tilde{x} )$ which results in  admissible $\Omega(d)$ values.

\section{Equilibrium Study} \label{sec:eq-setting}

The Nash equilibrium of a pricing strategy can be either a pure or mixed strategy. We prove that for a pure strategy, in the support of a mixed strategy, an $\epsilon$-Nash equilibrium exists. Consequently, multiple Nash equilibria can exist in this pricing game. Suppose a hypothetical equilibrium where a market price $\tilde{x}$ exists. We examine the hypothetical situation if one agent were to deviate from the $\tilde{x}$ of with a price of $x_n$. Particularly, we study the case when a player undercuts or prices above a set market price, where $x_n = \tilde{x} - d$ and $d$ is the value which the player deviates from the equilibrium price. From the equilibrium setting, as follows from Eq. \eqref{eq:taudes-demand-rewrite},

\begin{align} 
    f(\tilde{x}) &= \beta_0 + \beta_1 \tilde{x} + \beta_2(\tilde{x} - \bar{p}) \nonumber\\
    &= \beta_0 + \beta_1 \tilde{x} + \beta_2 \tilde{x} + c_N \label{eq:taudes-demand-rewrite} \\
    \text{where} \quad  c_N &= -\beta_2 \bar{p} \nonumber %\label{eq:cn-def}
\end{align}

We derive Eq. \eqref{eq:linear-demand-undercut} from Eq. \eqref{eq:taudes-demand-rewrite}. From Eq. \eqref{eq:linear-demand-undercut} we see that the expected demand is simply the demand function at equal pricing $f(\tilde{x})$ corrected by a factor of $\gamma d$ as defined in the Eq. \eqref{eq:gamma-def}, where $d$ is the amount the player $n$ deviates from the equilibrium price. % We stipulate $d > 0$ to specifically study the case of market undercutting. 

\begin{align} 
    f(x_n) &= \beta_0 + \beta_1 (x_n - d) + \beta_2(x_n - d - \bar{p}) \nonumber \\ 
    &= \beta_0 + \beta_1 x_n + \beta_2x_n + c_N - (\beta_1 + \beta_2) d \nonumber \\
    &= \beta_0 + \beta_1 x_n + \beta_2x_n + c_N + \gamma d \nonumber \\
    f(x_n) &= f(\tilde{x}) + \gamma d \label{eq:linear-demand-undercut}\\
    \text{where} \quad   x_n &\in \mathbb{R}, \gamma = - (\beta_1 + \beta_2), x > 0, d \in \mathbb{R} \label{eq:gamma-def}
\end{align}

% By extension from Conditions \eqref{eq:gamma-def}, $\gamma > 0$.

\subsection{Proof of a Best Response Function (Market Undercutting Scenario)} \label{sec:undercut}

\textbf{Proposition:} In an N-player game, under specific market conditions dictated by the reference price $\bar{p}$ and equilibrium price $\tilde{x}$, under certain proven conditions, stipulated later in Condition \ref{eq:undercut_conclusion}, there can exist a boundary $(0, d')$ such that the expected profit from deviating is greater than not deviating $\mathbf{E}[\Pi_n]^- > \mathbf{E}[\Pi_n]^0$, when deviating from the market price by undercutting with an amount of $d > 0$, as illustrated by Inequality \eqref{eq:eps-range-ded}.

% is always greater than not undercutting $\mathbf{E}[\Pi_n]^0$, that is $\mathbf{E}[\Pi_n]^- > \mathbf{E}[\Pi_n]^0$. 

\begin{align} 
    \frac{\mathbf{E}[\Pi_n]'}{\mathbf{E}[\Pi_n]^0} > 1, \quad \forall d \in (0, d') \label{eq:eps-range-ded}
\end{align}

Given the deviation constraint $d > 0$, therefore $\Phi_d > 1$, refer to Supplemental \ref{sec:win-prob}, we find the solution for the polynomial section of the gain function, defined by Eq. \eqref{eq:omega_p}.

\begin{align}
    \Omega (d)_p &= (1 + \frac{\gamma}{f(\tilde{x})} d) (1 - \frac{1}{\tilde{x}}d) \label{eq:omega_p} \\
    \Omega (d)_p &> 1 \label{eq:bound>1}
\end{align}

With the constraint $d > 0$, a solution to Inequality \eqref{eq:bound>1} is restricted by Condition \eqref{eq:omega_+_solution}. And such a solution only exists when $\tilde{x} > f(\tilde{x})/\gamma$.

\begin{align}
    0 < d &< \frac{ \frac{\gamma}{f(\tilde{x})} - \frac{1}{\tilde{x}} } { \frac{\gamma}{f(\tilde{x})\tilde{x}}} \label{eq:omega_+_solution} \\
    % 0 < d &< \frac{ \gamma \tilde{x} - f(\tilde{x})}{f(\tilde{x}) \tilde{x}}  \frac{f(\tilde{x})\tilde{x} }{\gamma} \\
    % 0 < d &< \frac{ \gamma \tilde{x} -  f(\tilde{x})}{\gamma} \\
    d &\in (0,  \tilde{x} -  f(\tilde{x})/\gamma) \quad \text{when} \quad  \tilde{x} >  f(\tilde{x})/\gamma \label{eq:d_undercut_bound}
\end{align}

Making substitutions from, Eq. \eqref{eq:taudes-demand-rewrite} into Eq. \eqref{eq:d_undercut_bound},

\begin{align}
    f(\tilde{x})/\gamma &< \tilde{x} \nonumber \\ %\label{eq:fxn_cond} 
    \frac{\beta_0 + \beta_1 \tilde{x} + \beta_2 \tilde{x} + c_N}{- (\beta_1 + \beta_2) } &< \tilde{x} \nonumber \\
    \frac{\beta_0 + c_N}{- (\beta_1 + \beta_2) } + \tilde{x} \frac{\beta_1 + \beta_2}{- (\beta_1 + \beta_2) } &< \tilde{x} \nonumber \nonumber \\
    \frac{\beta_0 + c_N}{- (\beta_1 + \beta_2) } &< \tilde{x} \Bigg( 1 + \frac{\beta_1 + \beta_2}{\beta_1 + \beta_2 } \Bigg) \nonumber \\
    - \frac{\beta_0 + c_N}{2 (\beta_1 + \beta_2) } &< \tilde{x} \nonumber \\
    \frac{\beta_2 \bar{p} - \beta_0 }{2 (\beta_1 + \beta_2) } &< \tilde{x}  \label{eq:xn_undercut_lowerbound}
\end{align}

We see from Inequality \eqref{eq:xn_undercut_lowerbound} that if the equilibrium price of the other agents $\tilde{x}$ is priced below a function of the reference price $\bar{p}$ there will be no value of $d$ that satisfies Inequality \eqref{eq:d_undercut_bound}. Therefore, the agent will not gain profit from undercutting if $\tilde{x}$, the current market price, is under a certain limit with respect a monotonic function the reference price $\bar{p}$. The exact conditions on which undercutting the $\tilde{x}$ will yield profit gain for any agent $n$ is outlined in Inequality \eqref{eq:undercut_conclusion}. 

% Making the substitutions from Eq. \eqref{eq:gamma-def},
% \begin{align}
%     0 < d &<  \frac{ -\beta_1 \tilde{x} - \beta_2 \tilde{x} - (beta_1 \tilde{x}   )  }{\gamma} 
% \end{align}

\begin{align}
     \frac{\beta_2 \bar{p} - \beta_0 }{2 (\beta_1 + \beta_2) } < \tilde{x}  \And d &\in (0,  \tilde{x} -  f(\tilde{x})/\gamma) \implies \mathbf{E}[\Pi_n]^- > \mathbf{E}[\Pi_n]^0 \label{eq:undercut_conclusion} 
\end{align}

$\square$

\subsection{Proof of the Existence of Multiple $\epsilon$-Nash Equilibria Single Stage Game}

\textbf{Proposition:} Suppose the market is in an equilibrium state where all agents price their items at a fixed price, and one player elects to undercut the market. We demonstrate that there is a theoretical maximum expected reward in this oligopoly for a single stage game for undercutting the market.

\begin{align}
    \mathbf{E}[\Pi_n]^- \leq \mathbf{E}[\Pi_n]^0 + \epsilon \label{eq:nash-def}
\end{align}

Inequality \eqref{eq:nash-def} expresses the conditions of an $\epsilon$-Nash equilibrium, that is no player can obtain a higher reward than a margin of $\epsilon$ by deviating from the equilibrium price of the market $\mathbf{E}[\Pi_n]^0$. In effect we acknowledge that the upper bound of $\Phi_d$ in Eq. \eqref{eq:softmax-gain} as,

\begin{align} 
    e^{ad} \frac{N_{\Phi} + e^{1- a\tilde{x}}}{N_{\Phi} + e^{1-a\tilde{x} + d}} &\leq e^{N_\phi ad} \\
    \Phi_d &\leq e^{N_\phi ad}
\end{align}

We take the partial derivative with respect the deviation amount $d$ to obtain theoretical maximum of $\Omega$.

\begin{align}
    \frac{\partial \mathbf{E}[\Pi_n]^-}{\partial d} &= \frac{\partial }{\partial d} \Bigg[e^{N_\phi ad} (1 + \frac{\gamma}{f(\tilde{x})} d) (1 - \frac{1}{\tilde{x}} d)\Bigg] \nonumber \\
    \frac{\partial \mathbf{E}[\Pi_n]^-}{\partial d} &= e^{N_\phi a} \frac{\partial }{\partial d} \Bigg[e^d (1 + \frac{\gamma}{f(\tilde{x})} d) (1 - \frac{1}{\tilde{x}} d)\Bigg]
\end{align}

Solve the derivative,

\begin{align}
    \frac{\partial \mathbf{E}[\Pi_n]^-}{\partial d} &= 0 \nonumber\\
    e^{N_\phi a} \frac{\partial }{\partial d} \Bigg[e^d (1 + \frac{\gamma}{f(\tilde{x})} d) (1 - \frac{1}{\tilde{x}} d)\Bigg] &= 0 \nonumber\\
    \frac{\partial }{\partial d} \Bigg[e^d + \frac{\gamma}{f(\tilde{x})} de^d - \frac{1}{\tilde{x}} de^d - \frac{\gamma}{f(\tilde{x})\tilde{x}} d^2e^d \Bigg] &= 0 \nonumber\\
    \frac{\partial }{\partial d} \Bigg[e^d + \Bigg( \frac{\gamma}{f(\tilde{x})} - \frac{1}{\tilde{x}} \Bigg) de^d - \frac{\gamma}{f(\tilde{x})\tilde{x}} d^2e^d \Bigg] &= 0 \nonumber\\
    e^d \Bigg[ \Bigg( \frac{\gamma}{f(\tilde{x})} - \frac{1}{\tilde{x}} \Bigg) (d + 1) - \frac{\gamma}{f(\tilde{x})\tilde{x}} d (d+2) + 1 \Bigg] &= 0 \nonumber\\
    \Bigg( \frac{\gamma}{f(\tilde{x})} - \frac{1}{\tilde{x}} \Bigg) (d + 1) - \frac{\gamma}{f(\tilde{x})\tilde{x}} d (d+2) + 1 &= 0 \label{eq:deriv-opt} 
\end{align}

The solution to Eq. \eqref{eq:deriv-opt}

\begin{align}
    d^* &= \frac{ \sqrt{c_1^2 - c_1 + 4(c_2-1)c_2 - 2c_2} }{2c_2}  \label{eq:d-star} \\
    \text{where} \quad  c_1 &= \frac{\gamma}{f(\tilde{x})} - \frac{1}{\tilde{x}}, \quad c_2 = \frac{\gamma}{f(\tilde{x})\tilde{x}}
\end{align}

We proved that when a player deviates from the market price by a margin of $d$, there exists an optimal deviation amount $d^*$, outlined in Eq. \eqref{eq:d-star} such that the expected profit gain $\Omega$ is maximized. $\epsilon$ is therefore,

\begin{align}
    \epsilon &= \mathbf{E}[\Pi_n]^- - \mathbf{E}[\Pi]^0 \nonumber\\
    &= \Omega(d^*) \mathbf{E}[\Pi_n]^0 - \mathbf{E}[\Pi_n]^0 \nonumber\\
    &=  \mathbf{E}[\Pi_n]^0 [\Omega(d^*) - 1)]  \label{eq:eps-d}
\end{align}

Multiple solutions can exist where $d^* = 0$, to give a perfect Nash equilibria solution, as $c_1$ and $c_2$ are functions of $\tilde{x}$. The Nash equilibrium condition is,

\begin{align}
    c_1^2 + 4(c_2-1)c_2 = c_1 + 2c_2 \quad  \xrightarrow[]{} \quad \epsilon = 0  \label{eq:ne-cond}
\end{align}

Therefore, by undercutting the market with any price $d > 0$, a player can theoretically yield no higher expected profits than $\epsilon$ greater than its competitors as defined in Eq. \eqref{eq:d-star} and Eq. \eqref{eq:eps-d}. Suppose a policy of pure strategy exists, $\pi^p_n$, where,

\begin{align} 
  \pi^p_n = & 
    \begin{cases} 
      1 &\text{for \quad} x = \tilde{x} \\ 
      0 &\text{for \quad} else
    \end{cases} \label{eq:pure-policy}
\end{align} 

Thus, $\pi^p_n$ constitutes an $\epsilon$-Nash equilibrium resulting from a pure strategy, with $\epsilon$ denoted in Eq. \eqref{eq:eps-d}, from which varying the parameters of $c_1$ and $c_2$ in Eq. \eqref{eq:d-star}, multiple $\epsilon$-Nash or Nash equilibria exist. $\square$

% \subsection{Complexity of computing Nash Equilibria}
\subsection{Existence of Nash equilibrium for Markov Game} \label{sec:eps-delta-exis}

% \begin{align} 
%     % v(s, \pi ) &= \int_{x_1}^\infty \int_{x_2}^\infty ... \int_{x_N}^\infty \pi(s, x_1)\pi(s, x_2) ... \pi(s, x_N) Q(s, \bar{x}) \ dx_1 dx_2 ... \ dx_N  \label{eq:nash-computation} \\
%     v(s, \pi_n', \pi_{-n}^*) &\leq v(s, \pi_n^*, \pi_{-n}^*) + \delta_N, \quad \forall s \in S \label{eq:nash-value-cond}
% \end{align} 

% To find a Nash equilibrium, one must propose a policy, and check if any possible deviation from the Nash policy $\pi^*$ is bounded by the value of $v(s, \pi^*)$ plus some threshold $\delta_N$ over all states. Ineq. \eqref{eq:nash-cond-v} presents a condition in which the algorithm to obtain a solution to is unlikely to be NP-Complete \citep{Daskalakis:2006}, and requires computation over all state and policy deviations.

\textbf{Proposition:} The existence of an $\epsilon$-Nash equilibrium in single stage game  ensures that there exists an $\delta$-Nash equilibrium in the Markov Game, as defined in Eq. \eqref{eq:nash-cond-v}, for the value of a joint policy regardless of the state transition behaviour of the Markov Game.

\textbf{Proof:} From \cite{filar:1997}, the value of a policy can be defined,

\begin{align} 
  \mathbf{v}(\pi) = \sum_{t=0}^{\infty} \gamma \mathbf{P}^t(\pi) \mathbf{R}(\pi) \label{eq:val-policy-inf}
\end{align} 

With a specific policy, the transition matrix of the Markov Game is known, and therefore the value of a policy can be expressed in Eq. \eqref{eq:val-policy-inf}, with defined transition $\mathbf{P}^t(\pi)$ and reward $\mathbf{R}(\pi)$ matrices. Provided the Nash equilibrium condition from Eq. \eqref{eq:nash-cond-v} we must demonstrate that,

\begin{align} 
  \mathbf{R}(\pi') \leq \mathbf{R}(\pi^*) + \mathbf{\epsilon} \xrightarrow[]{} \mathbf{V}(\pi') \leq \mathbf{V}(\pi^*) + \delta \label{eq:nash-v-proof}
\end{align} 

Where $\pi'$ denotes any joint policy, and $\pi^*$ denotes a Nash equilibrium policy. We know that the Markov transition probability holds such that,

\begin{align} 
  \sum_{i = 1}^{|S|}  \mathbf{P}^t(\pi)_{s i} = 1, \quad \forall s \label{eq:markov-trans-prop}
\end{align} 

$\mathbf{P}^t(\pi)_{s i}$ represents the probability of transition from state $s$ to state $i$. Eq. \eqref{eq:markov-trans-prop} simply indicates the sum of transition probabilities from state $s$ to any other state must equal 1 (Markov property). Furthermore, suppose,

\begin{align} 
  R_{s} &= \max_{1 \leq s < |S|} \{\mathbf{R}(\pi^*)_{s} + \epsilon \}  \label{eq:value-max} \\
  \delta_s &= R_{s} - \mathbf{P}(\pi^*)_s \times \mathbf{R}(\pi^*)
\end{align} 

$R_s$ represents the maximum reward obtainable from state $s$ under policy $\pi^*$, as illustrated in Eq. \eqref{eq:value-max}. Provided the Markov property on the transition matrix $\mathbf{P}^t(\pi)$ outlined in Eq. \eqref{eq:markov-trans-prop}, this implies that

\begin{align} 
  \mathbf{P}(\pi')_s \ \times \ \mathbf{R}(\pi') \leq \mathbf{P}(\pi^*)_s \ \times \ \mathbf{R}(\pi^*) + \delta_s \quad \forall s \in S
\end{align} 

Where $\delta$ is a maximum fixed value added to the value of the $\epsilon$-Nash equilibrium policy $\mathbf{V}(\pi^*)$ effectively bounding the value of any other policy deviating from $\pi^*$. $\square$

\clearpage

\onecolumn
\section{Plots and Figures} \label{sec:plots-figs}

\begin{figure}[H]
\minipage{0.5\textwidth}
  \includegraphics[width=\linewidth]{plots/market3.png}
  \caption*{\textbf{Market Scenario 1:} \\ $\beta_0 = 25, \beta_1 = -0.6, \beta_2 = -6.1, a = 0.1$}%\label{fig:market2}
\endminipage\hfill
\minipage{0.5\textwidth}
  \includegraphics[width=\linewidth]{plots/market2.png}
  \caption*{\textbf{Market Scenario 2:} \\ $\beta_0 = 15, \beta_1 = -1.05, \beta_2 = -3.1, a = 0.1$}%\label{fig:market3}
\endminipage\hfill
\minipage{0.5\textwidth}
  \includegraphics[width=\linewidth]{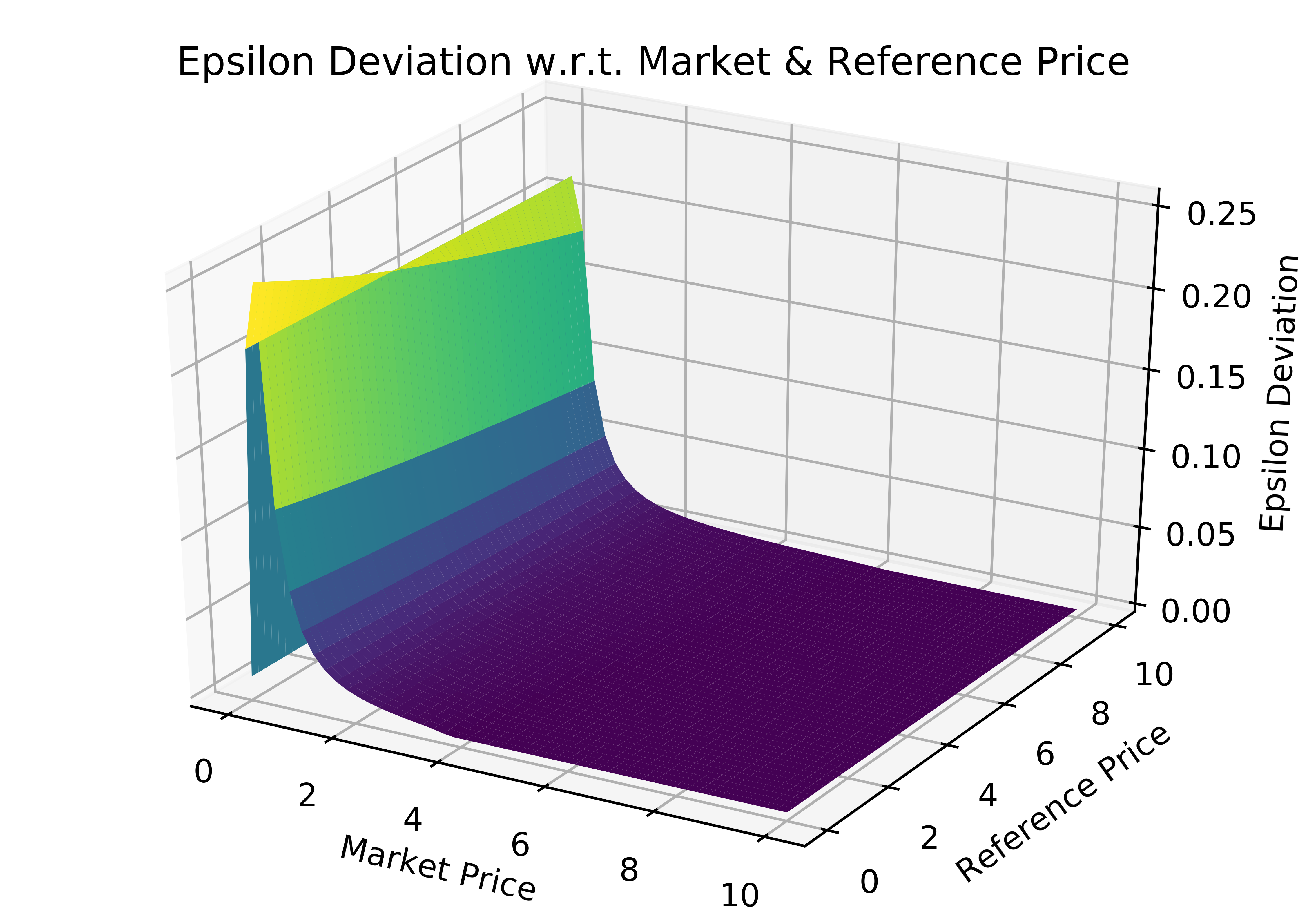}
  \caption*{\textbf{Market Scenario 3:} \\ $\beta_0 = 27, \beta_1 = -1.1, \beta_2 = -1.0, a = 0.1$}%\label{fig:market1}

\endminipage\hfill
\minipage{0.5\textwidth}
  \includegraphics[width=\linewidth]{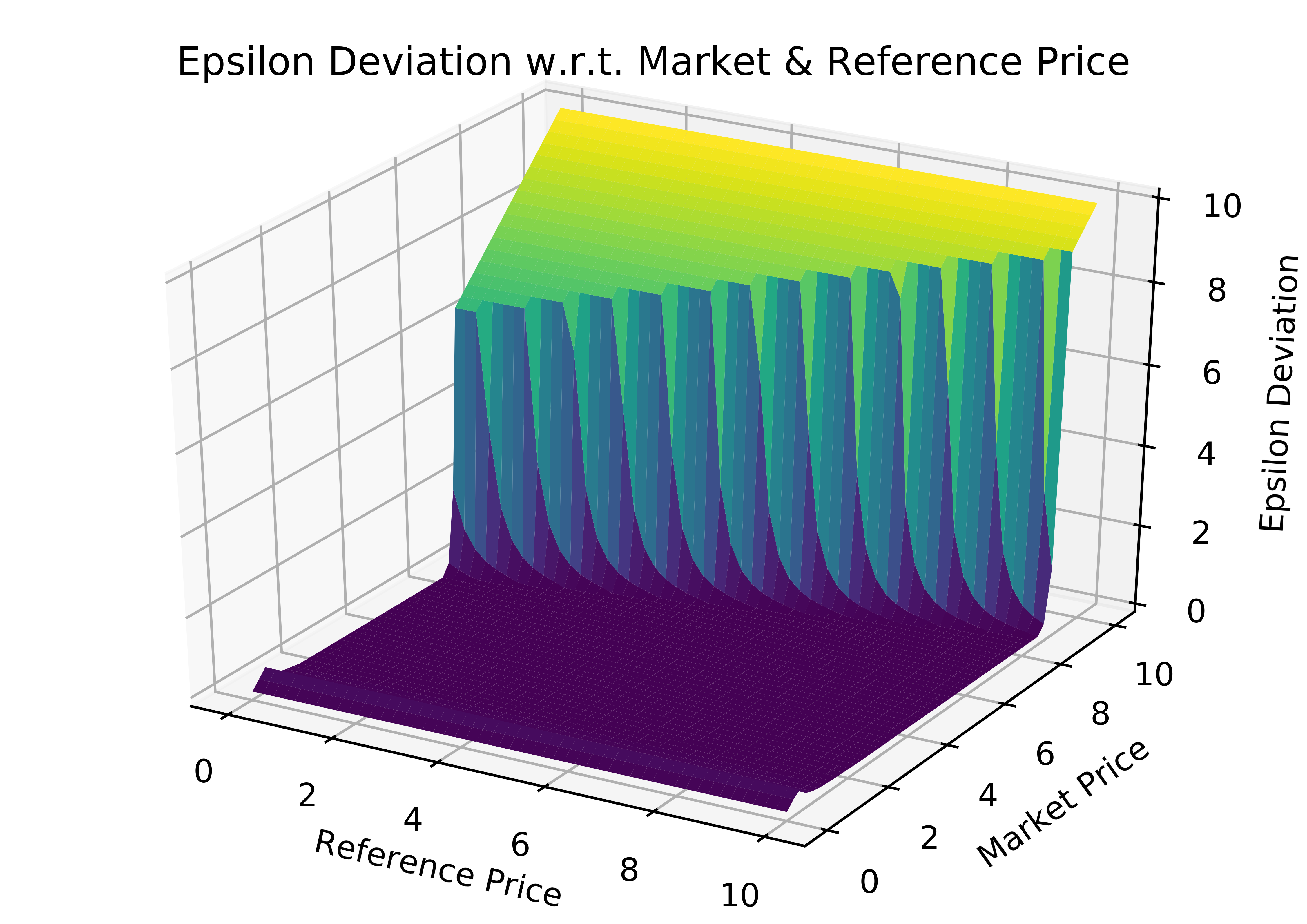}
  \caption*{\textbf{Market Scenario 4:} \\ $\beta_0 = 27, \beta_1 = -3.05, \beta_2 = -1.1, a = 0.2$}%\label{fig:market4}
\endminipage\hfill
\caption{Surface plot of maximum theoretical advantage from deviation ($\epsilon$-deviation) from Nash equilibrium with respect to market price $\tilde{x}$, and reference price $\bar{p}$, with their respective market parameters $\beta_0, \beta_1, \beta_2, a$ for arrival of a single sales event. The lower plateau area, illustrated in dark-blue, constitutes a state where the combination of joint action market price $\tilde{x}$, and reference price $\bar{p}$ do not result in significant difference when the player chooses to undercut the market. For each reference, there is a range of actions that result in $\epsilon$-Nash equilibrium, where $\epsilon < 0.0001$.} \label{fig:market-topo}
\end{figure}

\begin{figure}
\minipage{0.5\textwidth}
  \includegraphics[width=\linewidth]{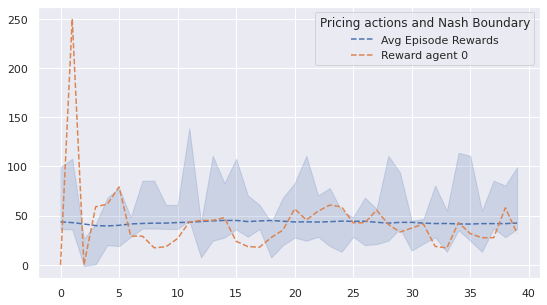}
  \caption*{Market rewards per episode Scenario 1 with $n=3$.}%\label{fig:market3_reward}
\endminipage\hfill
\minipage{0.5\textwidth}
  \includegraphics[width=\linewidth]{plots/market2_rewards.png}
  \caption*{Market rewards per episode Scenario 2 with $n=3$.}%\label{fig:market2_reward}
\endminipage\hfill
\minipage{0.5\textwidth}
  \includegraphics[width=\linewidth]{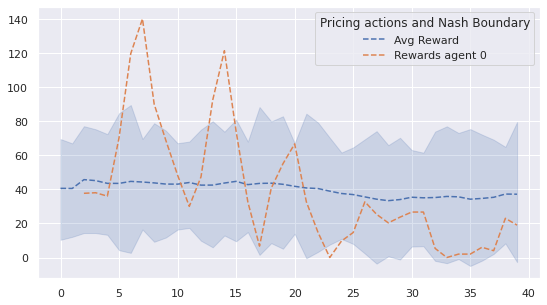}  
  \caption*{Market rewards per episode Scenario 3 with $n=3$.}%\label{fig:market1_bound}
\endminipage\hfill
\minipage{0.5\textwidth}
  \includegraphics[width=\linewidth]{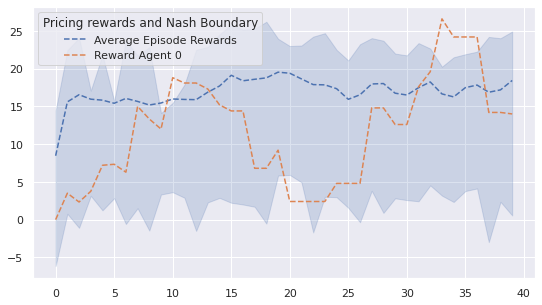}
  \caption*{Market rewards per episode Scenario 4 with $n=3$.}%\label{fig:market4_reward}
\endminipage\hfill
\minipage{0.5\textwidth}
  \includegraphics[width=\linewidth]{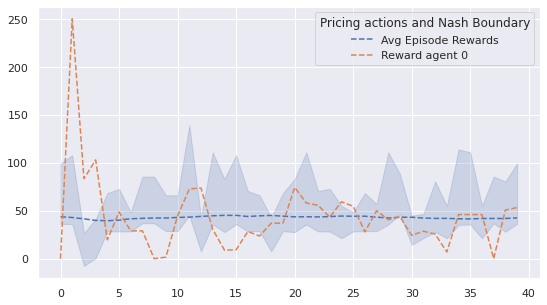}  
  \caption*{Market rewards per episode Scenario 2 with $n=5$.}%\label{fig:market3_bound}
\endminipage\hfill
\minipage{0.5\textwidth}
  \includegraphics[width=\linewidth]{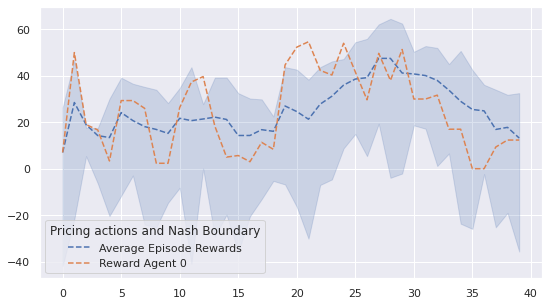}
  \caption*{Market rewards per episode Scenario 3 with $n=5$.}%\label{fig:market4_reward}
\endminipage\hfill
\caption{Average market rewards per episode overlaid on top of theoretical Nash equilibrium bounds (blue shade). Market average is in blue, and average reward of single agent in orange. In a Nash equilibrium, both the market average reward, and single agent reward should fall within the Nash equilibria boundary, as the MG progresses.}\label{fig:reward_plots}
\end{figure}

\begin{figure}
\minipage{0.5\textwidth}
  \includegraphics[width=\linewidth]{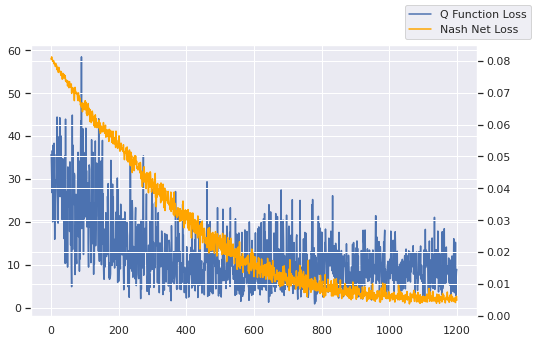}
  \caption*{Loss behaviour of Scenario 1 with $n=3$.}% \label{fig:market3_loss}
\endminipage\hfill
\minipage{0.5\textwidth}
  \includegraphics[width=\linewidth]{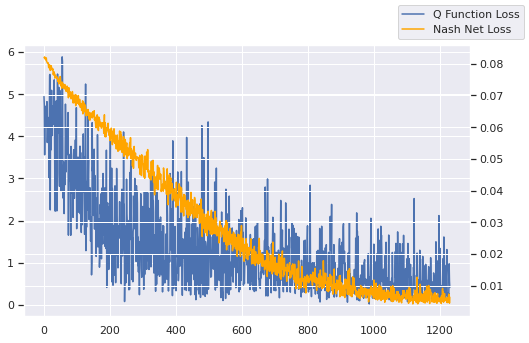}
  \caption*{Loss behaviour of Scenario 2 with $n=3$.}% \label{fig:market2_loss}
\endminipage\hfill
\minipage{0.5\textwidth}
  \includegraphics[width=\linewidth]{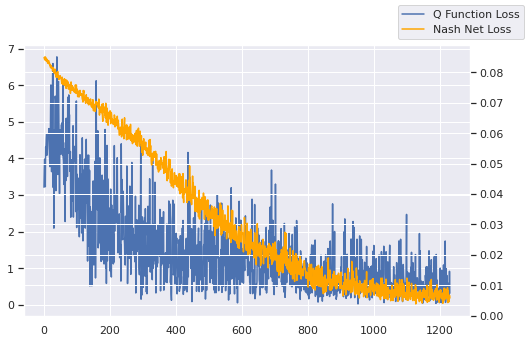}
  \caption*{Loss behaviour of Scenario 3 with $n=3$.}%\label{fig:market1_loss}
\endminipage\hfill
\minipage{0.5\textwidth}
  \includegraphics[width=\linewidth]{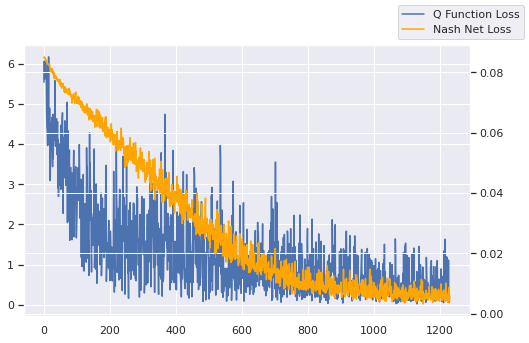}
  \caption*{Loss behaviour of Scenario 4 with $n=3$.}%\label{fig:market4_loss}
\endminipage\hfill
\minipage{0.5\textwidth}
  \includegraphics[width=\linewidth]{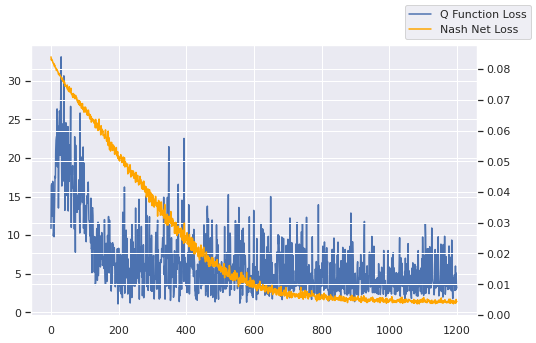}
  \caption*{Loss behaviour of Scenario 2 with $n=5$.}%\label{fig:market4_loss}
\endminipage\hfill
\minipage{0.5\textwidth}
  \includegraphics[width=\linewidth]{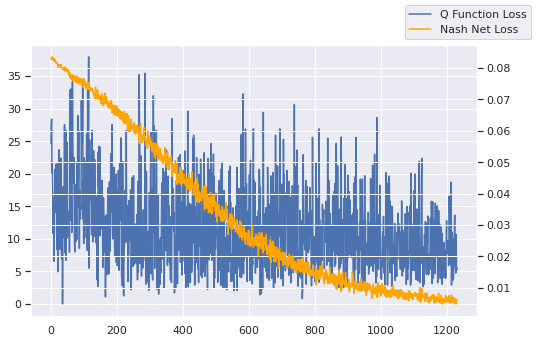}
  \caption*{Loss behaviour of Scenario 4 with $n=5$.}%\label{fig:market4_loss}
\endminipage\hfill
\caption{Loss behviour of batch update during Deep Q Learning (left y axis) and Nash Net update $\Psi(s)$ (right y axis).} \label{fig:loss_plot}
\end{figure}

% Because a this pure strategy $\pi^n_p$ is in the support of the $\epsilon$-Nash strategy, $\pi^{n^*}$, as long as $\pi(s, x_N) > 0$, 

% \begin{align}
%     v(\pi^n_p) \leq v(\pi^*) + \epsilon  \label{eq:eps-d}
% \end{align}

\clearpage

\end{document}